\documentclass[fleqn,usenatbib]{mnras}

\usepackage{newtxtext,newtxmath}

\usepackage[T1]{fontenc}

\DeclareRobustCommand{\VAN}[3]{#2}
\let\VANthebibliography\thebibliography
\def\thebibliography{\DeclareRobustCommand{\VAN}[3]{##3}\VANthebibliography}

\usepackage{graphicx}	
\usepackage{amsmath}	
\usepackage{amssymb}	
\usepackage{caption}
\usepackage{subcaption}
\usepackage{float}
\usepackage[verbose]{placeins}

\title[The TOI-1260 system]{Discovery of TOI-1260d and the characterisation of the multi-planet system\thanks{This article uses data from CHEOPS program CH\_PR100031.}}

\author[K. W. F. Lam et al.]{
Kristine W. F. Lam,$^1$\thanks{E-mail: kristine.lam@dlr.de} 
J. Cabrera,$^{1}$       
M.~J. Hooton,$^{2,3}$   
Y. Alibert,$^{2}$       
A. Bonfanti,$^{4}$      
M. Beck,$^{5}$          
A. Deline,$^{5}$        
\newauthor
H.-G. Flor{\'e}n,$^{6}$ 
A.~E. Simon,$^{2}$      
L. Fossati,$^{4}$       
C.~M. Persson,$^{7}$    
M. Fridlund,$^{7,8}$    
S. Salmon,$^{5}$          
S. Hoyer,$^{9}$         
\newauthor
H.~P. Osborn,$^{10,11}$ 
T~.G. Wilson,$^{12}$    
I.~Y. Georgieva,$^{7}$  
Gr. Nowak,$^{13,14}$    
R. Luque,$^{51}$        
J.~A. Egger,$^{2}$      
\newauthor
V. Abidekyan$^{15,16}$           
R. Alonso,$^{13,14}$             
G. Anglada Escud{\'e},$^{17,18}$ 
T. B\'{a}rczy,$^{19}$
D. Barrado,$^{20}$               
S.~C.~C. Barros,$^{15,16}$       
\newauthor
W. Baumjohann,$^{4}$             
T. Beck,$^{2}$                   
A. Bekkelien,$^{5}$              
W. Benz,$^{2}$                   
N. Billot,$^{5}$                 
X. Bonfils,$^{21}$               
A. Brandeker,$^{6}$              
\newauthor
C. Broeg,$^{2,10}$               
S. Charnoz,$^{22}$               
A. Collier Cameron,$^{12}$       
Sz. Csizmadia,$^{1}$             
M. B. Davies,$^{23}$             
M. Deleuil,$^{9}$                
\newauthor
L. Delrez,$^{24,25}$             
O.~D.~S. Demangeon,$^{15,16}$    
B.-O. Demory,$^{10}$             
D. Ehrenreich,$^{5}$             
A. Erikson,$^{1}$
A. Fortier,$^{2}$                
\newauthor
D. Futyan,$^{5}$
D. Gandolfi,$^{26}$              
M. Gillon,$^{24}$                
M. Guedel,$^{27}$
P. Guterman,$^{9,28}$
J. Laskar,$^{29}$                
D.~W. Latham, $^{30}$            
\newauthor
A. Lecavelier des Etangs,$^{31}$ 
M. Lendl,$^{5}$                  
C. Lovis,$^{4}$                  
K. Heng,$^{10,32}$               
K.~G. Isaak,$^{33}$
L. Kiss,$^{34,35,36}$
\newauthor
D. Magrin,$^{37}$                
P.~F.~L. Maxted,$^{38}$          
V. Nascimbeni,$^{39}$            
G. Olofsson,$^{6}$               
R. Ottensamer,$^{27}$
I. Pagano,$^{39}$                
\newauthor
E. Pall{\'e},$^{13,14}$          
G. Peter,$^{40}$                 
G. Piotto,$^{37}$                
D. Pollacco,$^{32}$
D. Queloz,$^{41,3}$              
I. Ribas,$^{17,18}$              
R. Ragazzoni,$^{37,42}$          
\newauthor
N. Rando,$^{43}$
H. Rauer,$^{1,44,45}$            
N.~C. Santos,$^{15,16}$          
G. Scandariato,$^{39}$           
S. Seager,$^{11,46,47}$
D. S{\'e}gransan,$^{5}$          
\newauthor
L.~M. Serrano,$^{26}$            
A.~M.~S. Smith,$^{1}$            
S.~G. Sousa,$^{15}$              
M. Steller,$^{4}$                
Gy.~M. Szab{\'o},$^{35,48}$      
N. Thomas,$^{2}$
\newauthor
S. Udry,$^{5}$                   
V. Van Grootel,$^{25}$           
N.~A. Walton,$^{49}$
J.~N. Winn.$^{50}$                
\\ \\
\textit{(Affiliations can be found after the references)}
}

\date{Accepted XXX. Received YYY; in original form ZZZ}

\pubyear{2022}

\begin{document}
\label{firstpage}
\pagerange{\pageref{firstpage}--\pageref{lastpage}}
\maketitle

\begin{abstract}
We report the discovery of a third planet transiting the star TOI-1260, previously known to host two transiting sub-Neptune planets with orbital periods of 3.127 and 7.493 days, respectively.
The nature of the third transiting planet with a 16.6-day orbit is supported by ground-based follow-up observations, including time-series photometry, high-angular resolution images, spectroscopy, and archival imagery.
Precise photometric monitoring with CHEOPS allows to improve the constraints on the parameters of the system, improving our knowledge on their composition.
The improved radii of TOI-1260b, TOI-1260c are $2.36 \pm 0.06 \rm R_{\oplus}$, $2.82 \pm 0.08 \rm R_{\oplus}$, respectively while the newly discovered third planet has a radius of $3.09 \pm 0.09 \rm R_{\oplus}$.
The radius uncertainties are in the range of 3\%, allowing a precise interpretation of the interior structure of the three planets. Our planet interior composition model suggests that all three planets in the TOI-1260 system contains some fraction of gas. The innermost planet TOI-1260b has most likely lost all of its primordial hydrogen-dominated envelope. Planets c and d were also likely to have experienced significant loss of atmospheric through escape, but to a lesser extent compared to planet b.

\end{abstract}




\section{Introduction}

Precise characterization of the bulk properties of transiting extrasolar planets allows constraining their possible interior composition.
This information is used to infer planet formation processes, as it can be used to demonstrate, for example, transport of material in the protoplanetary disk.
Additionally, planets orbiting close to their stars suffer from atmospheric erosion processes \citep[see, e.g.][]{2021A&A...648L...7L} that further shape their chemical evolution.
The CHaracterising ExOPlanet Satellite (CHEOPS) was launched in 2019 to allow the precise characterization of known planetary systems in order to better understand the processes of planetary formation and evolution~\citep{2021ExA....51..109B}.
Since the end of commissioning activities in April~2020, CHEOPS has successfully characterised several planetary systems \citep[e.g.][]{2020A&A...643A..94L,2022A&A...658A..75H}, including the discovery of new planets\citep[e.g.][]{Leleu2021,2021NatAs...5..775D}, improving our knowledge of planetary sciences.

In this paper we report the discovery of a third planet orbiting the system TOI-1260, which was previously known to host two planets \citep[][Hereafter G21]{2021MNRAS.505.4684G}.
The nature of the third planet is supported by ground-based follow-up observations, including time-series photometry, high-angular resolution images, spectroscopy, and archival imagery.
Precise photometric monitoring with CHEOPS allows to improve the constraints on the parameters of the system, improving our knowledge on their possible composition.
In particular, the study of multiplanet systems with sub-Neptune or super-Earths planets is very interesting for planet formation models, as they share the same disk and have evolved in the same timescales, yet with different outcomes~\citep[e.g.][]{2019ApJ...879...26K}.
The study of small planets allows exploring the effect of physical processes resulting in the observed variation of core compositions and envelope sizes~\citep{2020ApJ...891..158M}. 
Furthermore, multiplanetary systems provide excellent opportunity to study the dependence of planet formation, evolution and habitability on factors such as stellar insolation, age and spectral type~\citep[e.g.][]{2018AJ....155...48W,2018AJ....156..254W,Leleu2021}.

The planetary system around TOI-1260 was first discovered with the Transiting Exoplanet Survey Satellite \citep[TESS;][]{2014SPIE.9143E..20R}, a space-borne NASA mission launched in 2018 to survey the sky for transiting exoplanets around nearby and bright stars.
It builds on the legacy of the NASA’s Kepler space telescope \citep{2010Sci...327..977B} launched in 2009, which was the first exoplanet mission to perform a large statistical survey of transiting exoplanets. 
%
%
One of the goals of the TESS prime mission is to discover 50 exoplanets with radii smaller than $4R_{\oplus}$~(e.g. \citealt{2020Natur.583...39A,2021NatAs...5..775D,2021Sci...374.1271L}; and see also the overview of the planet yield during the Prime Mission in~\citealt{2021ApJS..254...39G}). 
Coordinated mass measurements via precise high-resolution spectroscopic follow-up enable accurate inferences about the bulk composition and atmospheric characterization of small exoplanets.
To date, there are more than 100 exoplanets smaller than $4R_{\oplus}$ in the public domain, with many more in the TESS pipeline.

CHEOPS and TESS missions complement each other in their aims, with TESS carrying the weight of the detection efforts, organizing the community for the ground-based support observations, and CHEOPS providing accurate measurements of the planetary radius, allowing detailed characterization of the planetary interiors~\citep[e.g.][]{2022MNRAS.511.4551L,2022MNRAS.511.1043W}.

The paper is structured as follows. 
Section~\ref{TESS} describes the TESS observations and transit analysis. 
The CHEOPS observations and its transit analysis is described in Section~\ref{sec:CHEOPSLC}. 
The HARPS-N data and the spectral analyses are described in Section~\ref{sec:HARPSN}. 
Section~\ref{sec:joint_fit} outlines the the model and result of the joint analysis of the TESS, CHEOPS photometry and HARPS-N RVs. 
Section~\ref{sec:discussion} discusses the results of the global fit, the interior structure of the planets, planet atmospheric evolution model and the possible origin of the planetary system. 
Finally, the conclusion of our work is presented in Section~\ref{sec:conclusion}.

\section{TESS Photometry}
\label{TESS}
\begin{figure*}
	\includegraphics[width=2\columnwidth]{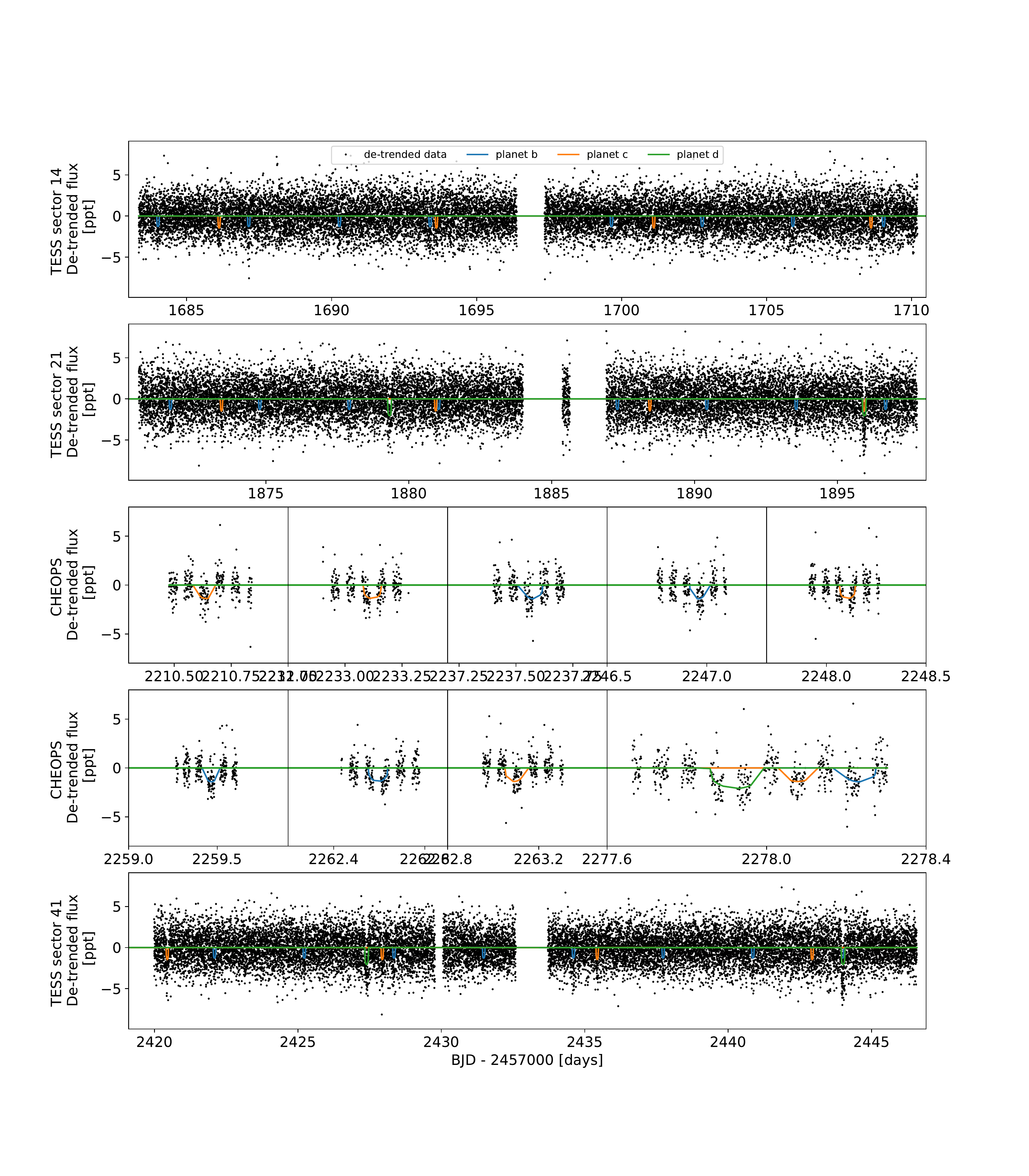}
    \caption{Time-series light curves of TOI-1260. From top to bottom: The TESS PDCSAP light curves from Sectors 14, 21, and 41 are shown in the first, second, and last panels, respectively. The TESS light curves were detrended using a Gaussian Process model described in Section~\ref{sec:joint_fit}. The CHEOPS light curves are shown in the third and fourth rows.}
    \label{fig:TESS_lc}
\end{figure*}

TOI-1260 was observed by TESS during sector 14 (between 18 Jul 2019 and 15 Aug 2019 on camera 4, CCD 3) and sector 21 (between 21 Jan 2020 and 18 Feb 2020 on camera 2, CCD 2) in 2-minute short cadence mode.
This data set was previously analyzed in G21.
The target was further observed in sector 41 (between 23 Jul 2021 and 20 Aug 2021) in 2-minute and 20-second cadence mode. 
The TESS data were process by the Science Process Operation Centre \citep[SPOC][]{2010SPIE.7740E..23T,2017ksci.rept....6M}. 
SPOC extracted TESS light curves using a Simple Aperture Photometry (SAP) and known instrumental systematics are corrected in the Presearch Data Conditioning (PDCSAP) light curves \citep{2012PASP..124.1000S,2012PASP..124..985S,2014PASP..126..100S}. 
The TESS PDCSAP light curves were downloaded from Mikulski Archive for Space Telescopes (MAST\footnote{\url{https://archive.stsci.edu/tess}}) and were used for subsequent analyses.
Figure \ref{fig:TESS_lc} shows the PDCSAP light curves of TOI-1260.


\section{CHEOPS Photometry}
\label{sec:CHEOPSLC}
We performed follow-up photometric observations with CHEOPS to refine the radii of the two inner planets and to confirm the presence of the outer planet, scheduling 9 visits between 26 Dec 2020 and 4 Mar 2021.

The discovery paper of TOI-1260b and c (G21) reported a possible third planet, on the basis of a single transit in sector 21, with a number of period aliases in the range 20.3 d < P < 56.3 d.
The paper discussed the possibility of the third planet having a period of 16.6 days. At the time, only one clear single transit were observed in the TESS light curves. The 16.6-day signal in the radial velocity (RV) data was not significant due to the period being close to a harmonic of the stellar rotation period. The nature of the 16.6-day signal was uncertain. 
However, their results encouraged our efforts to confirm the suspected third planet in the system.
We used our code to identify possible additional transit signatures in the existing data~\citep[the code is described in][]{2022arXiv220303194O}.
This identified a unique period of 16.6 d, meaning that the transit fell in the gap in sector 14. 
The available data at that time was used to constrain the possible ephemeris of the putative third planet in the system.
The visits that we programmed with CHEOPS lasted between 8.25 and 16.8 hours to cover the transits of planet b and c, as well as to confirm the presence of the third planet candidate. 
The details of each observation runs are listed in Table \ref{tab:CHEOPS_obs}. 

Observations obtained in each visit were processed by the CHEOPS data reduction pipeline \citep[DRP;][]{2020A&A...635A..24H}. 
The pipeline calibrated each image by applying bias, gain, non-linear effects, dark current, and flat field corrections. 
It also corrects individual calibrated frames from environmental effects such as smearing trails, bad pixels, background, and stray-light pollution. 
The DRP then performed aperture photometry on the calibrated and corrected images to extract the photometric fluxes. 
Next, the DRP pipeline provides four sets of light curves by performing aperture photometry on the calibrated images using different aperture sizes ($R_{\rm ap}$). These apertures are RINF ($R_{\rm ap} = 22.5"$), DEFAULT ($R_{\rm ap} = 25"$), RSUP ($R_{\rm ap} = 30"$), and a further aperture OPTIMAL which is optimised for each visit. 
We used the root-mean-squared (RMS) values of the light curve extracted by different aperture in each visit to assess the the light curves. 
Apart from the first visit of planet c, the RINF aperture of each visit gives the lowest RMS. Thus the corresponding light curves were use for subsequent analysis. For the first visit of planet c, we used the light curve reduced from the OPTIMAL aperture for subsequent analysis.

It is known that the rotation of the CHEOPS field-of-view along with the orbit of the spacecraft can result in varying background, contaminants, or other non-astronomical sources ~\citep[e.g.][]{2022MNRAS.511.1043W}. This may induce noises in the data and cause short trm trends in the photometric light curve. Fortunately, the DRP pipeline provides basis vectors for CHEOPS which is used to correct and detrend these variabilities in the light curves. For our dataset, we use the open-source Python package ~\texttt{pycheops} \citep{2021MNRAS.tmp.3057M} to evaluate the data produced by DRP and found that the light curves showed periodic flux variation that is in phase with the orbit of the spacecraft. 

For each visit, we performed simultaneous transit fitting and detrending of a combinations of standard basis vectors used in the decorrelation of CHEOPS data (i.e. background, contamination, smear, x and y centroid positions, and first, second, and third-order harmonics of the roll angle). The Bayesian Information Criterion (BIC) and minimum $\chi^2$ of the model in each visit were assessed separately to select the basis vectors required to optimally detrend each set of light curve. We also used the \texttt{$add_glint$} function to remove internal reflection from resulting from the spacecraft rotation cycle in each visit. The detrended CHEOPS light curves were used for our joint model described in Section~\ref{sec:joint_fit}.

\begin{table*}
\center
\caption{List of CHEOPS observations of TOI-1260. The file key is the unique identifier which corresponds to the dataset used. \label{tab:CHEOPS_obs}}
\begin{tabular}{lllccc}
\hline
File Key & Observation Start & Observation End & Duration {[}h{]} & Exposure Time {[}s{]} & $\rm N_{frames}$ \\
\hline
CH\_PR100031\_TG018501\_V0200 & 2020-12-26 23:21 & 2020-12-27 08:04  & 8.72  & 60.0 & 296 \\
CH\_PR100031\_TG018502\_V0200 & 2021-01-18 09:34 & 2021-01-28 18:34  & 9.00  & 60.0 & 270 \\
CH\_PR100031\_TG018503\_V0200 & 2021-02-02 09:51 & 2021-02-02 18:15  & 8.40  & 60.0 & 285 \\
CH\_PR100031\_TG018504\_V0200 & 2021-02-17 10:49 & 2021-02--17 19:14 & 8.42  & 60.0 & 291 \\
CH\_PR100031\_TG036501\_V0200 & 2021-01-22 21:31 & 2021-01-23 04:58  & 7.45  & 60.0 & 273 \\
CH\_PR100031\_TG036502\_V0200 & 2021-02-01 05:58 & 2021-02-01 14:13  & 8.25  & 60.0 & 254 \\
CH\_PR100031\_TG036504\_V0200 & 2021-02-13 18:18 & 2021-02-14 02:33  & 8.25  & 60.0 & 281 \\
CH\_PR100031\_TG036505\_V0200 & 2021-02-16 22:17 & 2021-02-17 06:32  & 8.25  & 60.0 & 280 \\
CH\_PR100031\_TG038201\_V0200 & 2021-03-04 03:50 & 2021-03-04 19:10  & 15.34 & 60.0 & 522 \\
\hline
\end{tabular}
\end{table*}

\section{Host star characterisation}
\label{sec:HARPSN}
 
TOI-1260 was observed between 14 Jan 2020 and 13 Jun 2020, a campaign in which 33 high resolution spectra (R$=115 000$) were reported by G21 using the HARPS-N spectrograph \citep[][]{2012SPIE.8446E..1VC}. 
The HARPS-N Data Reduction Software (DRS) pipeline \citep{2014SPIE.9147E..8CC} was used to extract the spectra. 

To retrieve the fundamental parameters of TOI-1260, stellar effective temperature, $T_\mathrm{eff}$, iron abundance relative to hydrogen, [Fe/H], and the  surface gravity, $\log g$, we modelled the HARPS-N co-added high resolution spectrum with the  spectral analysis package \href{http://www.stsci.edu/~valenti/sme.html}{{\tt{SME}}} \citep[Spectroscopy Made Easy;][]{1996A&AS..118..595V,2017A&A...597A..16P}, version 5.22. 
With atomic and molecular line data from \href{http://vald.astro.uu.se}{VALD}~\citep{Ryabchikova2015}, the MARCS 2012~\citep{Gustafsson08} atmosphere grids, and a chosen set of fundamental parameters, {\tt{SME}} calculate synthetic stellar spectra which is fitted to the observations. 
The models were also checked with the Atlas12~\citep{Kurucz2013} grids. 
We followed the modelling procedure explained in~\citep{2018A&A...618A..33P}. 
In summary, we modelled $T_\mathrm{eff}$ and $\log g$ with the H$_\alpha$ line wings and  the \ion{Ca}{I} $\lambda$=6102~\AA, 6122~\AA, and 6162~\AA~triplet, respectively. 
The model was checked with the \ion{Na}{I} doublet at $\lambda$=5888~\AA~and 5895~\AA. 
The abundances and  projected stellar rotational velocity, $V \sin i_\star$, were modelled from unblended lines between $\lambda$=6000~\AA~and 6600~\AA. 
The results, listed in Table~\ref{tab:star_param}, were checked with the empirical {\tt {SpecMatch-Emp}} code \citep{2018AJ....155..127H} which were in very good agreement with {\tt{SME}}. 
The full set of host star parameters are listed in Table \ref{tab:star_param}.

As recently described in \citet{Schanche2020}, we can use a modified version of the infrared flux method (IRFM; \citealt{Blackwell1977}) to determine the stellar angular diameters and effective temperatures of stars through known relationships between these properties, and estimates of the apparent bolometric flux, via a Markov-Chain Monte Carlo (MCMC) approach. 
We perform synthetic photometry of TOI-1260 by building spectral energy distributions (SEDs) from stellar atmospheric models with the stellar parameters, derived via the spectral analysis detailed above, as priors. 
To compute the apparent bolometric flux, these fluxes are compared to the observed data taken from the most recent data releases for the following bandpasses; {\it Gaia} G, G$_{\rm BP}$, and G$_{\rm RP}$, 2MASS J, H, and K, and {\it WISE} W1 and W2 \citep{Skrutskie2006,Wright2010,GaiaCollaboration2021} with the stellar atmospheric models taken from the \textsc{atlas} Catalogues \citep{Castelli2003}. 
We convert the stellar angular diameter to the stellar radius of TOI-1260 using the offset corrected {\it Gaia} EDR3 parallax \citep{Lindegren2021} and obtain $R_{\star}=0.672\pm0.010\, R_{\odot}$.

Together with $R_{\star}$, we used the effective temperature and the metallicity to then derive the isochronal mass $M_{\star}$ and age $t_{\star}$. 
Rather than directly adopting [Fe/H] as a proxy for the stellar metallicity, we estimated the $\alpha$-element abundance by averaging out the [Mg/H] and [Si/H], obtaining $[\alpha/\mathrm{Fe}]=0.13\pm0.13$. 
Using Eq. (3) from \citet{yi01}, we finally computed the metallic content of the star ($\mathrm{[M/H]}=0\pm0.15$ dex) from [Fe/H] and [$\alpha$/Fe].
To make our $M_{\star}$ and $t_{\star}$ estimates more robust we employed two different evolutionary models, namely PARSEC\footnote{\textit{PA}dova and T\textit{R}ieste \textit{S}tellar \textit{E}volutionary \textit{C}ode: \url{http://stev.oapd.inaf.it/cgi-bin/cmd}} v.1.2S \citep{marigo17} and CLES \citep[Code Liègeois d'Évolution Stellaire][]{scuflaire08}. 
In detail, we interpolated the input set ([M/H], $T_{\mathrm{eff}}$, and $R_{\star}$) within pre-computed grids of PARSEC isochrones and tracks through the isochrone placement technique described in \citet{bonfanti15,bonfanti16} and we derived a first best-fit pair of mass and age. 
The code further accounted for $v\sin{i}$ and $\log{R'_{\mathrm{HK}}}$ as outlined in \citet{bonfanti16} to improve the convergence. 
Instead, the second pair of mass and age was inferred by directly fitting the input set into the evolutionary track built by CLES according to the Levenberg-Marquadt minimisation criterion \citep{salmon21}.
After carefully checking the consistency of the results outputted by the two codes through the $\chi^2$-test described in \citet{bonfanti21}, we finally merged the respective output distributions ending up with $M_{\star}=0.679_{-0.057}^{+0.095}\,M_{\odot}$ and $t_{\star}=6.7_{-5.2}^{+5.1}$ Gyr. The host star mass and radius derived in this work are consistent within $\sim 1$-sigma and we adopt values from this work for subsequent analyses.

\begin{table*}
\caption{Stellar parameters of TOI-1260. \label{tab:star_param}}
\begin{tabular}{lll}
\hline
Parameter [Unit]                             & Value.               & Note \\
\hline
Identifiers                                  & TIC 355867695             &        \\
RA (ICRS Ep. 2016.0)                         & 157.14401106413           & 1      \\
Dec (ICRS Ep. 2016.0)                        & +65.85418726790           & 1      \\
$\pi$ [mas]                                  & 13.6226 $\pm$ 0.0147      & 1      \\
$\mu_{\alpha}$ [mas yr$^{-1}$]               & -177.340 $\pm$ 0.012      & 1      \\
$\mu_{\delta}$ [mas yr$^{-1}$]               & -81.693 $\pm$ 0.013       & 1      \\
\hline
Effective temperature $T_{\rm eff}$ [K]      & $4227 \pm 85$             & 2      \\
{[Fe/H]} abundance                           & $-0.1 \pm 0.07$           & 2      \\
{[Si/H]} abundance                           & $-0.02 \pm 0.15$          & 2      \\
{[Mg/H]} abundance                           & $0.09 \pm 0.15$          & 2      \\
{[$\alpha$/Fe]} abundance                    & $0.13\pm0.13$             & 2      \\
{[M/H]} abundance                            & $0\pm0.15$                & 2      \\
$\log g$ [cgs]                               & $4.57 \pm 0.05$           & 2      \\
Stellar rotation velocity $v\sin{i}$ [km s$^{-1}$] & 1.5 $\pm$ 0.7       & 2      \\
Stellar rotation period $P_{\mathrm{rot}}$ [d] & $30.63 \pm 3.81$       & 2      \\
Chromospheric activity $\log{R'_{\mathrm{HK}}}$ & $-4.86$                & 3      \\
Stellar mass $M_{\rm star}$ [$M_{\odot}$]    & $0.679_{-0.057}^{+0.095}$ & 2      \\
Stellar radius $R_{\rm star}$ [$R_{\odot}$]  & $0.672 \pm 0.010$         & 2      \\
Stellar density $\rho_{\rm star}$ [g cm$^{-3}$]   & $3.43 \pm 0.08$     & 2      \\
Bolometric luminosity [$L_{\odot}$]          & $0.129 \pm 0.004$         & 2      \\
Stellar age [Gyr]                            & $6.7_{-5.2}^{+5.1}$       & 2      \\
\hline
[1] \cite{GaiaCollaboration2021}, [2] this work, [3] \citet{suarez15}
\end{tabular}
\end{table*}

\section{Joint light curve and radial velocity analysis}
\label{sec:joint_fit}

A global analysis of the observational data was performed using the \texttt{exoplanet} toolkit~\citep{exoplanet:joss}. 
The toolkit implements the probabilistic programming package \texttt{PyMC3}~\citep{pymc3} to perform a Bayesian inference using a Hamiltonian Monte Carlo \citep[HMC;][]{1987PhLB..195..216D} method.

We first removed the out of transit variability in the TESS light curve by first masking the transits in the light curve, then binning the light curve into 1-hour steps. A Gaussian Process (GP) regression model with a simple harmonic oscillator (SHO) kernel, implemented by \texttt{celerite2}~\citep{celerite1,celerite2}, was then applied to remove the light curve variations.

The joint analysis was subsequently carried out on the "flattened" TESS light curve from the aforementioned best-fit GP photometry model, CHEOPS light curve, and the HARPS-N RV data. The toolkit uses \texttt{starry}~\citep{2019AJ....157...64L} to model the limb darkened transit light curves. To account for the limb darkening parameters of the star, we used the quadratic limb darkening coefficients ($u_1$, $u_2$) parameterised by \citet{2013MNRAS.435.2152K} in the model for each photometric instrument. Uniform priors were used for the planet orbital periods ($P_\mathrm{b}$,$P_\mathrm{c}$ and $P_\mathrm{d}$), mid-transit times ($T0_\mathrm{b}$, $T0_\mathrm{c}$ and $T0_\mathrm{d}$), planet-to-star radius ratios ($R_\mathrm{p, b}/R_\mathrm{star}$, $R_\mathrm{p, c}/R_\mathrm{star}$ and $R_\mathrm{p, d}/R_\mathrm{star}$) and impact parameters ($b_\mathrm{b}$, $b_\mathrm{c}$, $b_\mathrm{d}$). We account for the instrument zero-point offset between the TESS ($\sigma_\mathrm{TESS}$) and CHEOPS ($\sigma_\mathrm{CHEOPS}$) light curves by fitting a mean to the light curves of the two separate instruments. Gaussian priors were used for the stellar mass $M_\mathrm{star}$ and radius $R_\mathrm{star}$ based on the results in Section \ref{sec:HARPSN}. The Keplerian orbits of the three transiting planets are defined by their orbital periods. The planets' respective semi-major axes ($a_b$, $a_c$, $a_d$) can be derived using Kepler's third law and the scaled semi-major axes ($a_\mathrm{b}/R_\mathrm{star}$, $a_\mathrm{c}/R_\mathrm{star}$, $a_\mathrm{d}/R_\mathrm{star}$) were subsequently derived from the fitted stellar radius.

The TOI-1260 star is moderately active where activity-induced variations were reported by G21. The activity-induced variations in the RVs were modeled by a GP model alongside the three-planet Keplerian model. We chose a \texttt{RotationTerm} GP kernel \citep{celerite2}, which consists of a mixture of two SHO terms to describe the stellar rotation. A uniform prior was used for the log rotation period ($\log P_\mathrm{rot}$) parameter and the radial velocity semi-amplitudes ($K_\mathrm{b}$, $K_\mathrm{c}$, $K_\mathrm{c}$) in the RV dataset. Finally, we included jitter ($\sigma_\mathrm{HARPS}$) and mean velocity offset or systemic offset ($\gamma_\mathrm{HARPS}$) parameters for the RV fit.
The host star mass and radius were sampled using a Gaussian prior which is based on our results in Section~\ref{sec:HARPSN}. We note that the best-fit stellar rotation period from our GP model is $30.63 \pm 3.81$ days. This gives a rotation rate of $2\pi R_\mathrm{star} / P_\mathrm{rot} = 1.1~\mathrm{km~s^{-1}}$ which is consistent with our $V \sin{i}$ value from Section \ref{sec:HARPSN}.

The fitted parameters were first optimised with the \texttt{scipy.optimize.minimize} function, integrated in the \texttt{exoplanet} package, to find the respective maximum a posteriori parameters.
These estimates were used to initialise parameters in the sampling space via a “No U-Turn Sampling” \citep[NUTS;][]{2011arXiv1111.4246H}, a gradient-based HMC sampler implemented in \texttt{PyMC3}.
We initiated 4 sampling chains where each chain has 2000 tuning steps and 2000 draw iterations. The Gelman-Rubin statistic~\citep{1992StaSc...7..457G} of the sample is $\leq 1.003$, indicating the chains are converged. 

\begin{figure}
     \centering
     \begin{subfigure}[b]{\columnwidth}
         \centering
         \includegraphics[width=\columnwidth]{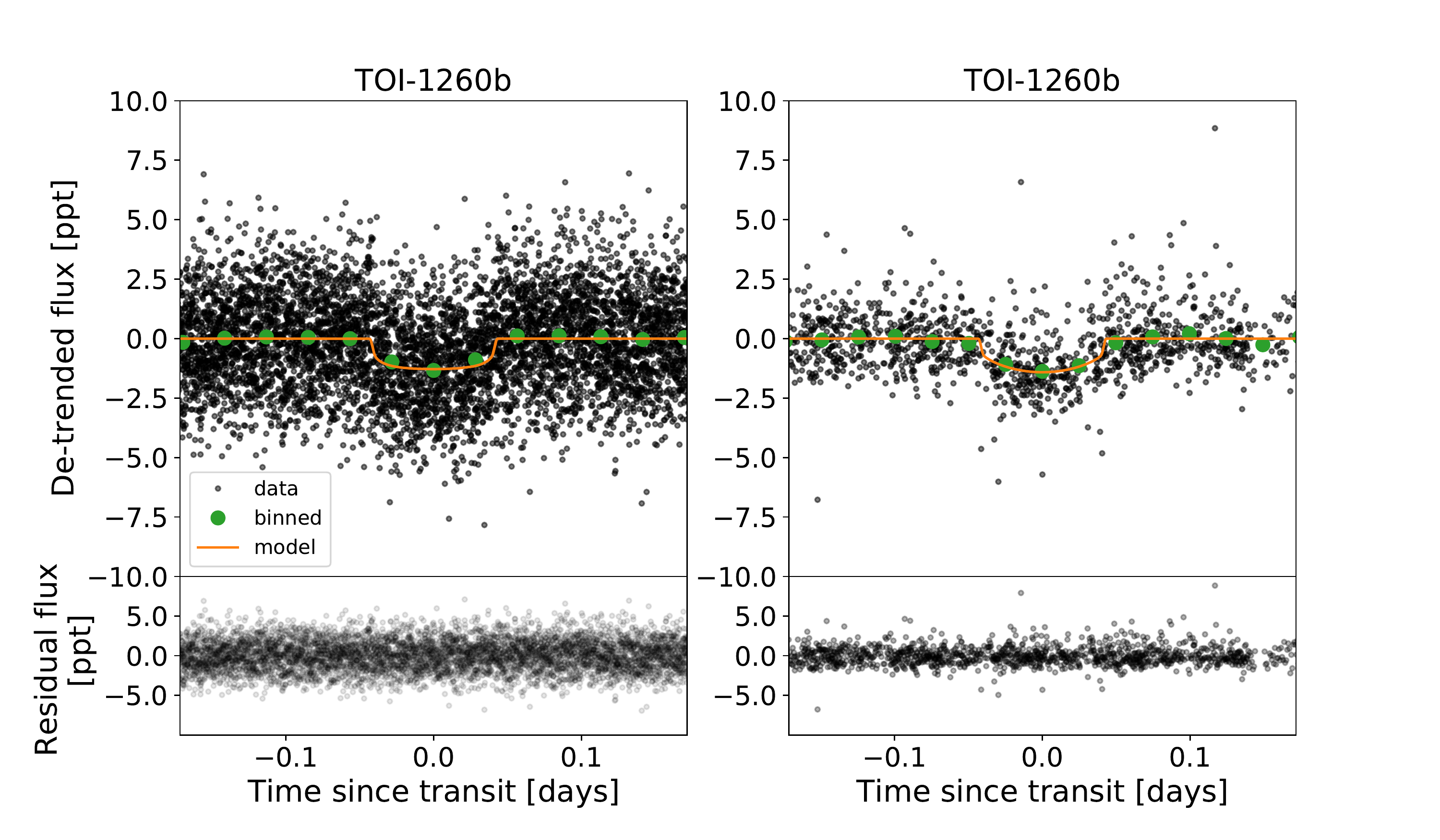}
     \end{subfigure}
     \hspace*{\fill}
     \newline
     \hfill
     \begin{subfigure}[b]{\columnwidth}
         \centering
         \includegraphics[width=\columnwidth]{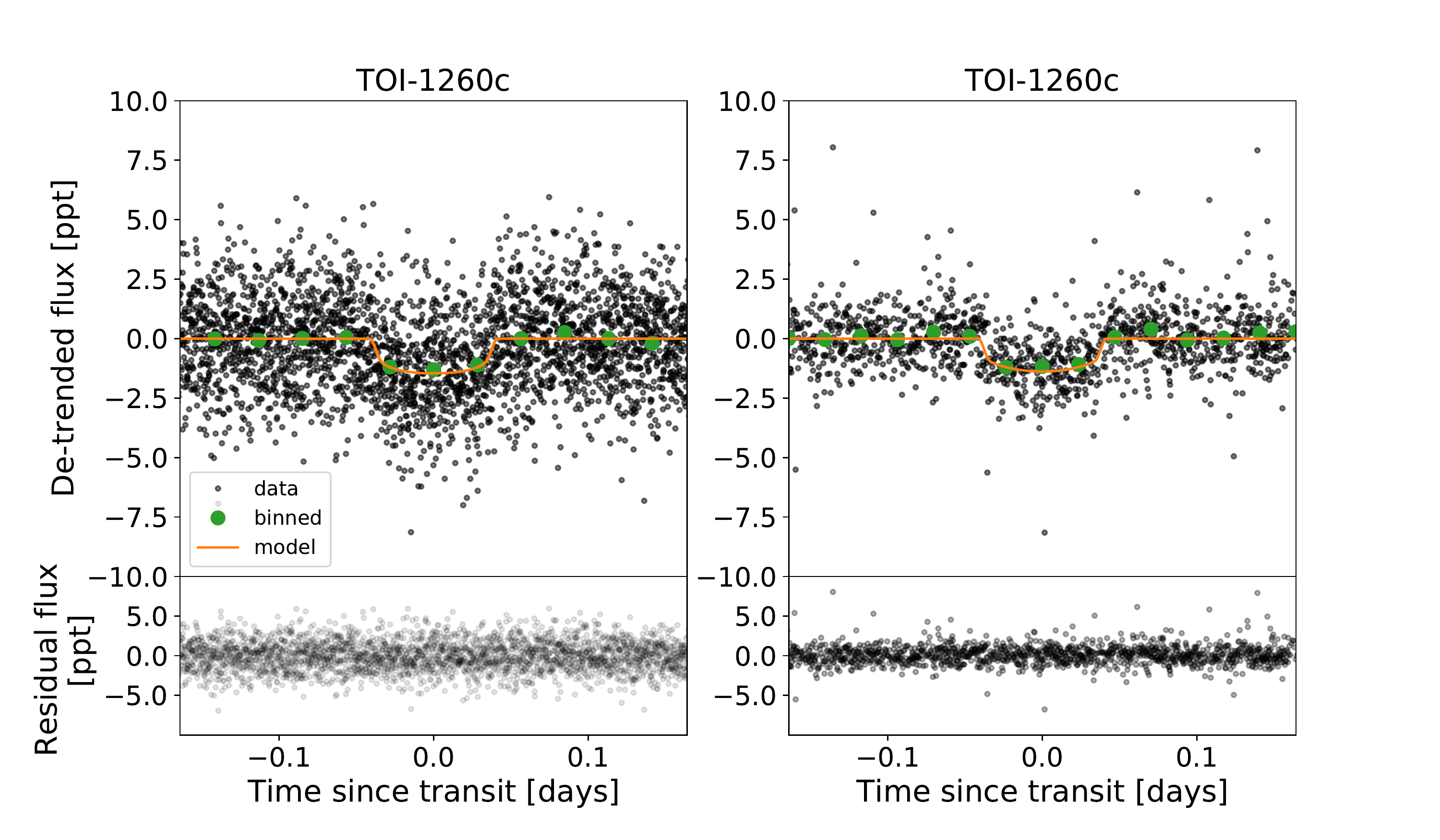}
     \end{subfigure}
     \hspace*{\fill}
     \newline
     \hfill
     \begin{subfigure}[b]{\columnwidth}
         \centering
         \includegraphics[width=\columnwidth]{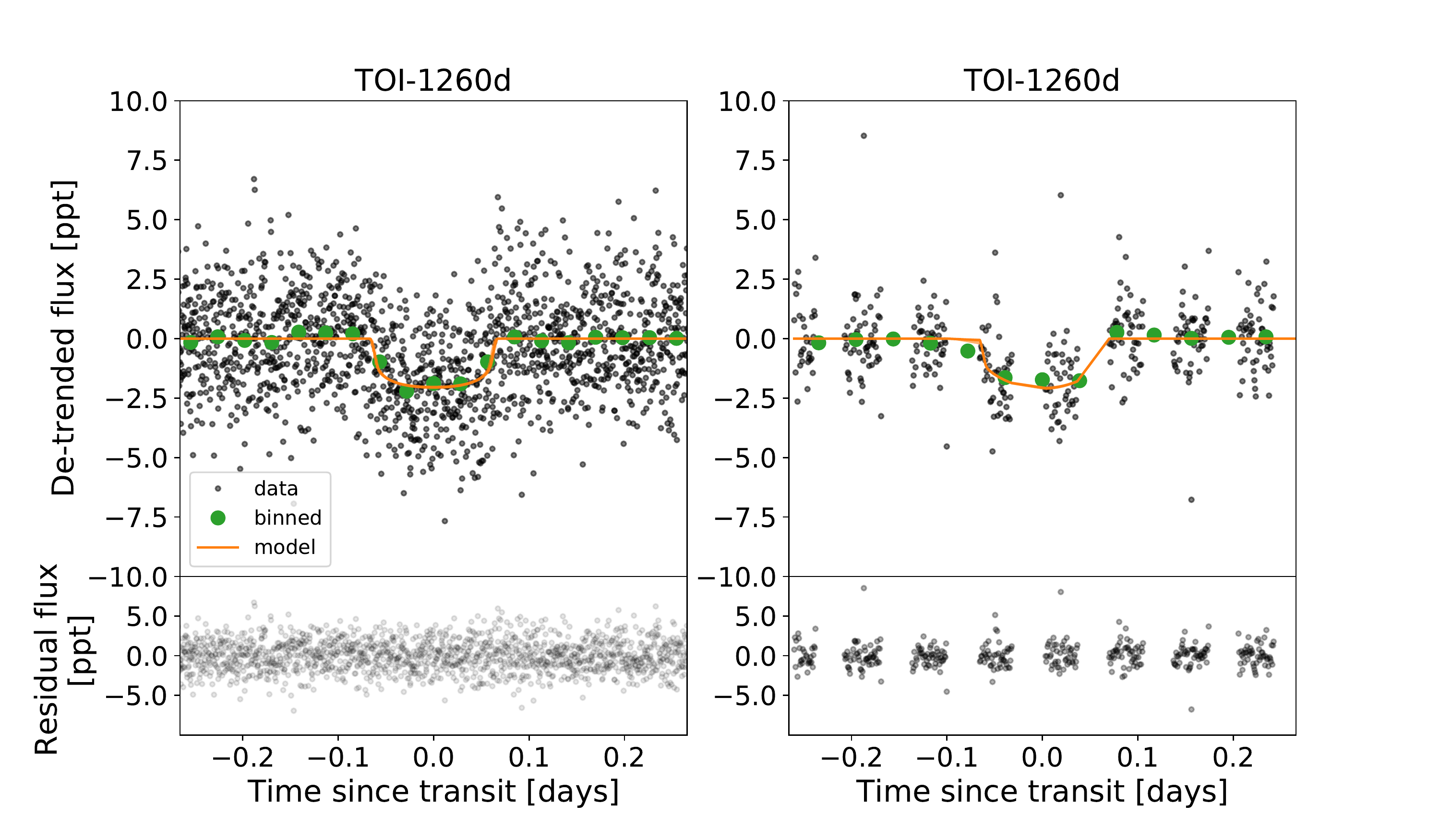}
     \end{subfigure}
     \hspace*{\fill}
     \centering
        \caption{Phase folded light curves of TOI-1260b (top), TOI-1260c (middle), TOI-1260d (bottom). The TESS data are shown in the left panels and the CHEOPS data are shown in the right panels. Residuals of each transit are shown below each phase-folded light curves. The phase binned data are denoted by green points and the orange line shows the best-fitted transit models for each planet.
        }
        \label{fig:TOI-1260_foldedLC}
\end{figure}

\begin{figure*}
	\includegraphics[width=2\columnwidth]{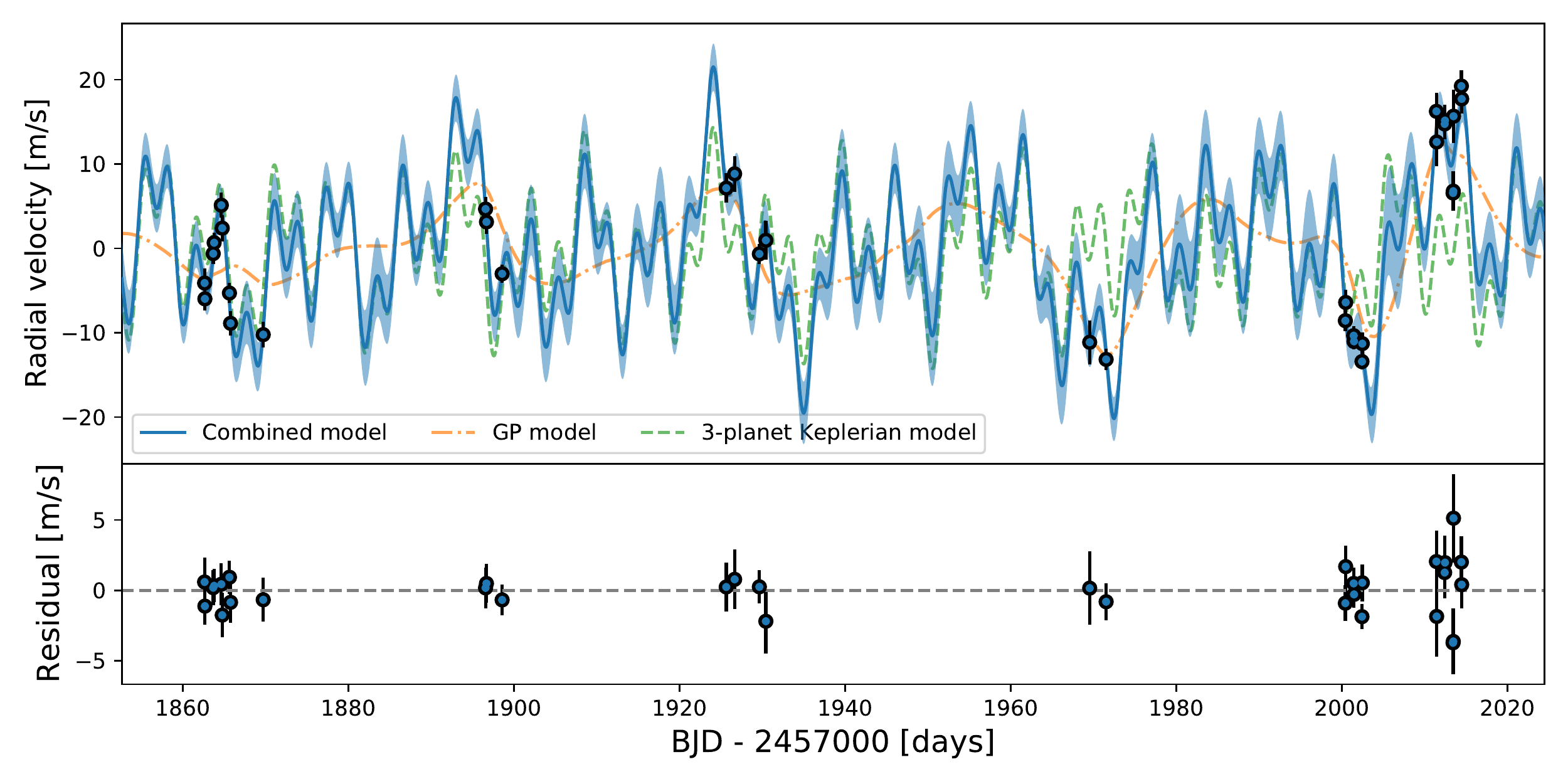}
    \caption{\textit{Top:} Time-series HARPS-N RVs of TOI-1260. The RVs were modelled using a three-planet Keplerian RV model and a GP simultaneously to model the activity-induced RV variations (see Section \ref{sec:joint_fit}). The green dash line shows the 3-planet Keplerian model and the orange dash-dot line shows the GP model that accounts for activity induced RV variations. The blue solid line shows the median three-planet Keplerian + GP model. The 1-sigma credible intervals of the best fit  Keplerian + GP model is indicated by the blue shaded region. }  \textit{Bottom:} Residuals of the RV data.
    \label{fig:HARPS_RV}
\end{figure*}

\begin{figure}
     \centering
     \begin{subfigure}[b]{\columnwidth}
         \centering
         \includegraphics[width=\columnwidth]{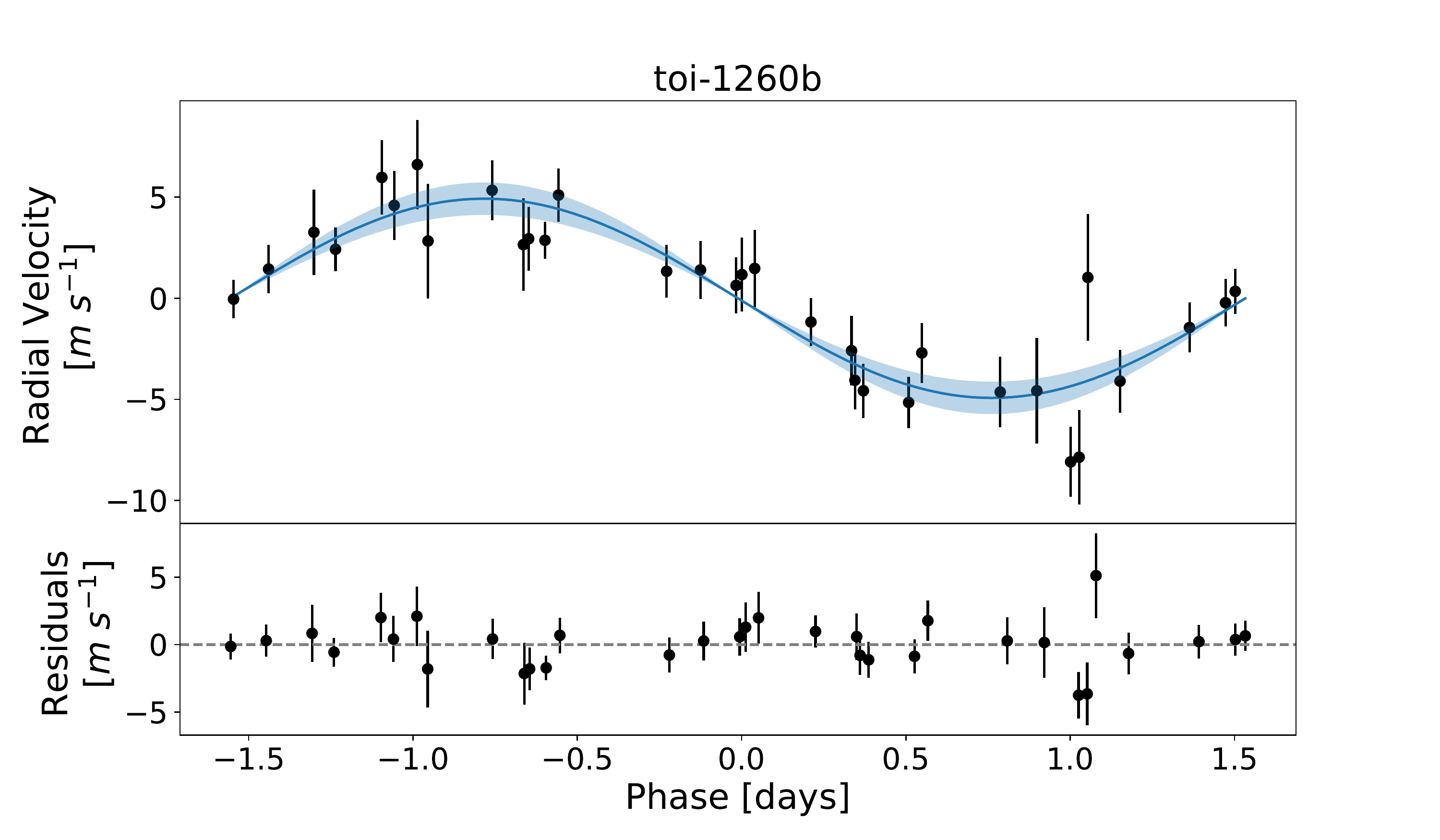}
     \end{subfigure}
     \hfill
     \begin{subfigure}[b]{\columnwidth}
         \centering
         \includegraphics[width=\columnwidth]{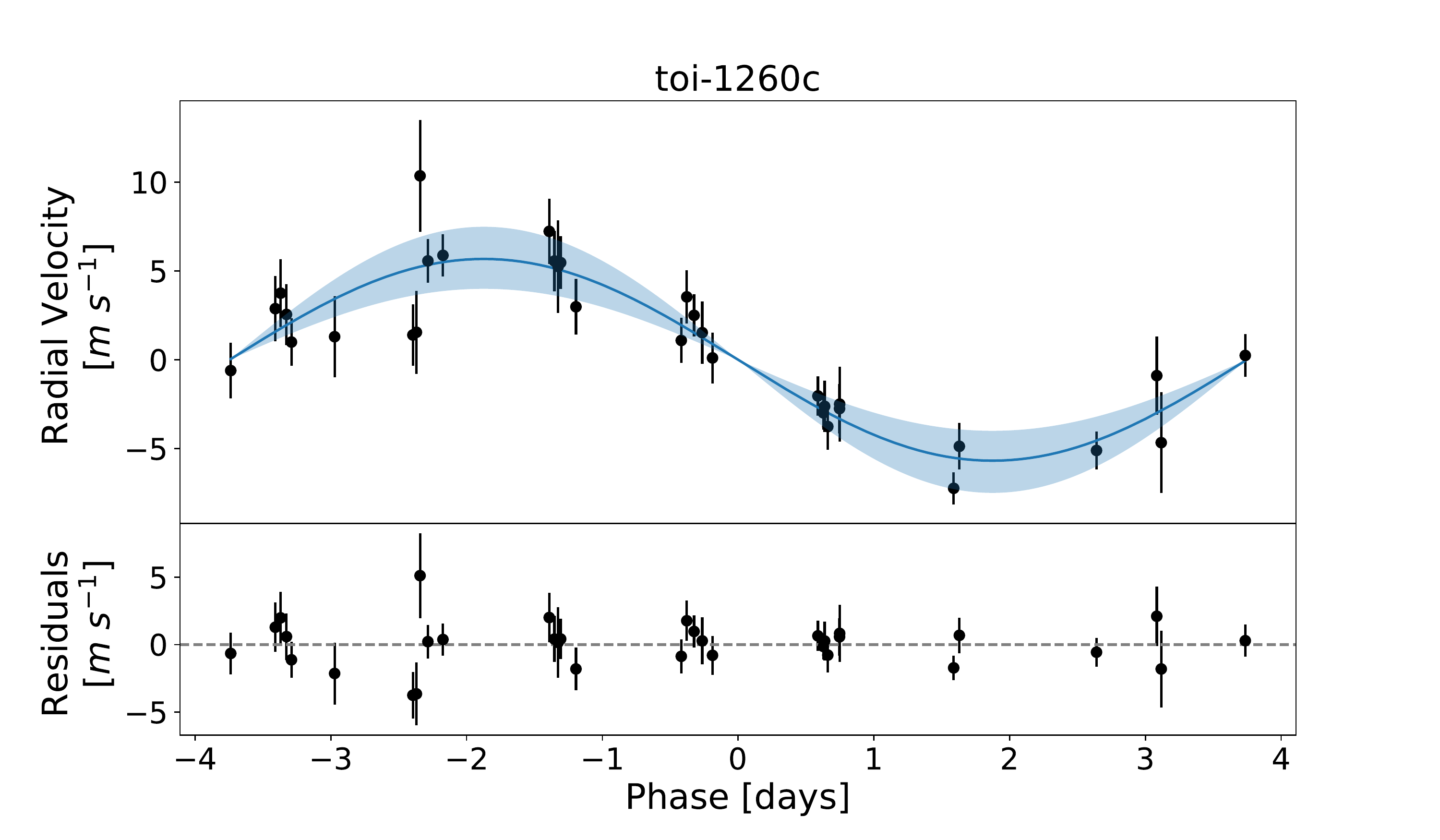}
     \end{subfigure}
     \hfill
     \begin{subfigure}[b]{\columnwidth}
         \centering
         \includegraphics[width=\columnwidth]{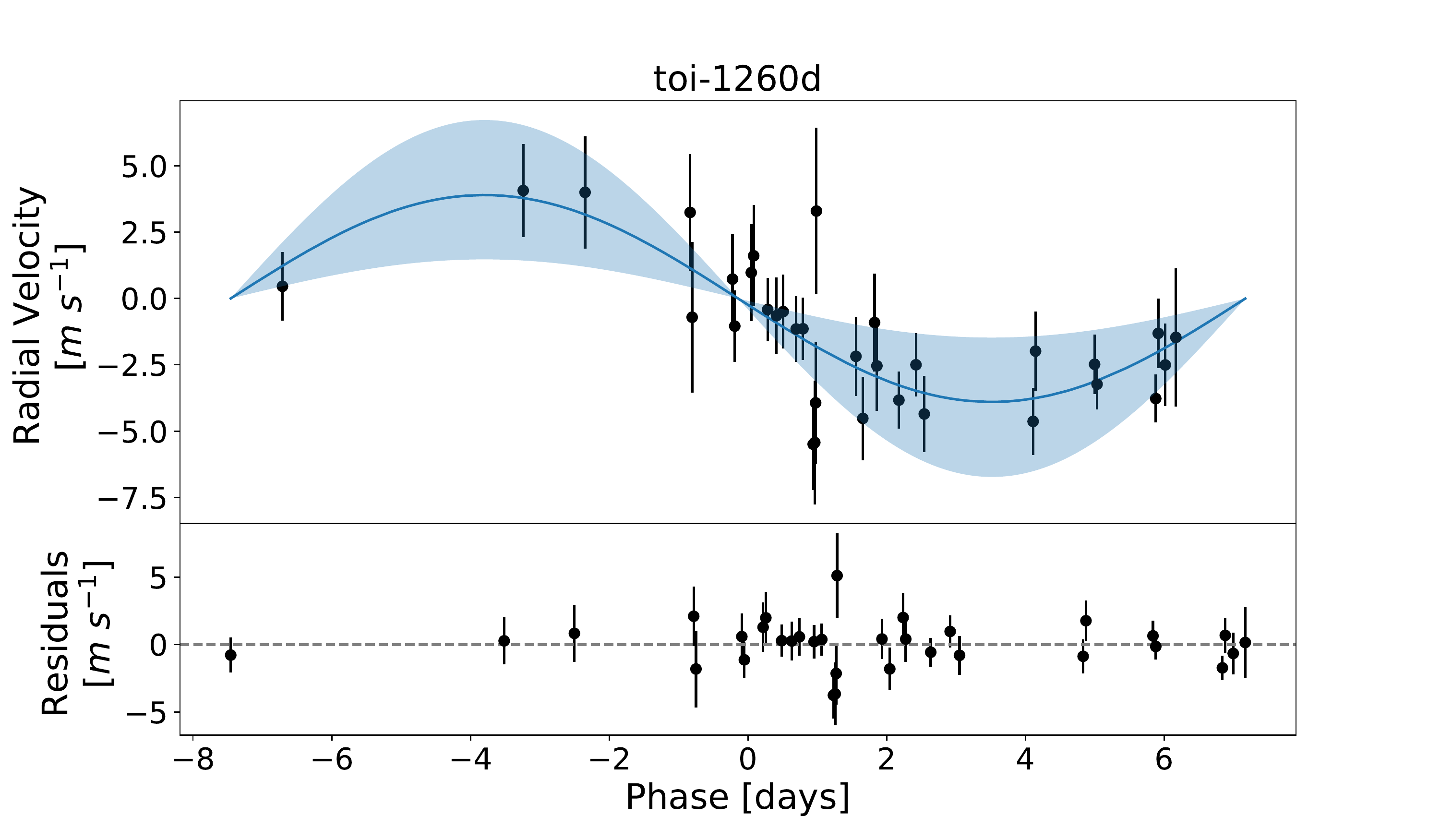}
     \end{subfigure}
        \caption{Phase-folded radial velocities and residuals of TOI-1260b (top), TOI-1260c (middle), TOI-1260d (bottom). The best-fit RV models are indicated by the solid blue line and the corresponding 1-sigma credible interval is shown by the blue shaded region.}
        \label{fig:TOI-1260_foldedRV}
\end{figure}

The phase-folded TESS and CHEOPS transit light curves and the corresponding best-fit transit models are shown in Figure~\ref{fig:TOI-1260_foldedLC}. The HARPS-N RVs and best-fit three planet RV model is shown in Figure~\ref{fig:HARPS_RV}. The phase-folded RVs for each planet and their respective best-fit models are shown in Figure~\ref{fig:TOI-1260_foldedRV}.

We studied the case where planet eccentricities are allowed to float in the model and found that there are no difference between the zero and non-zero eccentricities models. Hence we adopted the zero eccentricity model. The resulting median parameters and their 1-$\sigma$ uncertainties are listed in Table \ref{tab:system_param}. The posterior distributions of fitted parameters are shown in the corner plot in Figure \ref{fig:corner_plot}.

TOI-1260 is a multiplanet system that consists of three transiting exoplanets where the innermost planet TOI-1260b has a radius and mass of $2.41 \pm 0.05 ~R_{\mathrm \oplus}$ and $8.56 \pm 1.54 ~M_{\mathrm \oplus}$, respectively. TOI-1260c has a radius and mass of $2.74 \pm 0.07 ~R_{\mathrm \oplus}$ and $13.20 \pm 4.23 ~M_{\mathrm \oplus}$, respectively, while the outermost planet TOI-1260d has a radius of $3.12 \pm 0.08 ~R_{\mathrm \oplus}$ and a mass of $11.84 \pm 7.79 ~M_{\mathrm \oplus}$, respectively. 
With the addition of the CHEOPS photometry as well as TESS data from more recent sectors, this work has significantly improved the precision of the radius measurements of TOI-1260b and c compared to previous work. The radii of all three transiting planets are measured with a precision of better than 3\%.
We note that the mass precision of planets b and c in our work is does not improve despite the inclusion of planet d in the Keplerian model. This may be due to the methodology used to model the stellar activity induced variation in the RV data. In G21, the author applies a multi-dimensional GP approach and used activity indicators as prior to constraining the GP model which reduced the flexibility of the GP to model the RVs and may have resulted in a smaller semi-amplitude precision. Nevertheless, the mass determination of planets b and c are consistent within 1-sigma with values derived in G21.

The mass precision of the planets is the main source of uncertainty in the determination of the planetary bulk densities in the system. This work highlights the need to strategically obtain more RVs for the system in order to understand the effect of stellar activity on the RVs of the system and better constrain the planetary masses.

\newpage

\begin{table*}
\caption{System parameters obtained from the joint light curves and radial velocities analysis. The median values and 1-sigma uncertainty are reported. \label{tab:system_param}}
\begin{tabular}{lccc}
\hline
Parameter {[}Unit{]}                       & Planet b            & Planet c             & Planet d             \\
\hline
\multicolumn{2}{l}{\textit{Fitted parameters}}                          &                               &                           \\
Period P {[}day{]}                         & $3.127463 \pm 0.000005$    & $7.493134 \pm 0.000020$       & $16.608164 \pm 0.000083$  \\
Epoch T0 {[}BJD-2457000{]}                 & $2065.564269 \pm 0.000396$ & $2068.270505 \pm 0.000577$    & $2062.017406 \pm 0.001309$\\
Planet-to-Stellar radius ratio {[}Rp/Rs{]} & $0.0329 \pm 0.0006$        & $0.0377 \pm 0.0007$           & $0.0425 \pm 0.0009$       \\
Impact paramater b                         & $0.20 \pm 0.12$            & $0.75 \pm 0.02$               & $0.53 \pm 0.05$       \\
Radial velocity semi-amplitude K [m s$^{-1}$] & $4.93 \pm 0.83$          & $5.67 \pm 1.77$               & $3.90 \pm 2.54$           \\
Eccentricity e         & 0 (adopted)     & 0 (adopted)          & 0 (adopted)               \\
Angle of periastron $\omega$ [$^{\circ}$] &  0 (adopted)     &      0 (adopted)  &   0 (adopted)      \\
\multicolumn{4}{l}{\textit{Derived parameters}}                                                                                     \\
Transit duration T14 {[}hr{]}              & $2.06 \pm 0.02$            & $1.97  \pm 0.03$              & $3.19 \pm 0.07$           \\
Transit depth {[}ppm{]}                    & $1082 \pm 37$              & $1421 \pm 55$                 & $1808 \pm 78$             \\
Scaled semi-major axis a/Rs                & $11.73 \pm 0.35$           & $20.99 \pm 0.63$              & $35.69 \pm 1.06$          \\
Orbital semi-major axis a {[}au{]}         & $0.0367 \pm 0.0011$        & $0.0657 \pm 0.0020$           & $0.1116 \pm 0.0033$       \\
Inclination i {[}deg{]}                    & $89.03 \pm 0.61$           & $87.97 \pm 0.11$              & $89.14 \pm 0.10$        \\
Planet radius $R_p$ [$R_{\oplus}$]         & $2.41 \pm 0.05$            & $2.76 \pm 0.07$               & $3.12 \pm 0.08$           \\
Planet mass $M_p$ [$M_{\oplus}$]           & $8.56 \pm 1.54$            & $13.20 \pm 4.23$              & $11.84 \pm 7.79$          \\
Planet density $\rho_p$ [\rm $g~cm^{-3}$]  & $3.35 \pm 0.64$            & $3.45 \pm 1.14$               & $2.14 \pm 1.42$           \\
Planet surface gravity $\log g_p$          & $3.16 \pm 0.09$            & $3.23 \pm 0.15$               & $3.08 \pm 0.30$                     \\
Equilibrium dayside temperature {[}K{]}    & $871 \pm 24$               & $651 \pm 18$                  & $499 \pm 14$                 \\
Stellar insolation [$S_{\oplus}$]          & $95.58 \pm 0.07$           & $29.81 \pm 0.05$             & $ 10.32 \pm 0.07 $         \\
                                           &                            &                               &                      \\
TESS instrument offset $\sigma_\mathrm{TESS}$ [ppm]             & \multicolumn{3}{l}{$64.0 \pm 8.6$}                          \\
CHEOPS instrument offset $\sigma_\mathrm{TESS}$ [ppm]           & \multicolumn{3}{l}{$48.8 \pm 14.7$}                          \\
HARPS jitter $\sigma_\mathrm{HARPS}$ [m s$^{-1}$]                & \multicolumn{3}{l}{$0.22 +/- 0.79$}                          \\
Systemic radial velocity $\gamma_\mathrm{HARPS}$ [m s$^{-1}$]    & \multicolumn{3}{l}{$10.73 \pm 2.63$}                          \\
Limb darking parameter $u_{1,\mathrm{TESS}}$            & \multicolumn{3}{l}{$0.21 \pm 0.18$}                            \\
Limb darkening parameter $u_{2,\mathrm{TESS}}$          & \multicolumn{3}{l}{$0.53 \pm 0.26$}                            \\
Limb darking parameter $u_{1,\mathrm{CHEOPS}}$          & \multicolumn{3}{l}{$0.92 \pm 0.18$}                            \\
Limb darking parameter $u_{2,\mathrm{CHEOPS}}$          & \multicolumn{3}{l}{$-0.33 \pm 0.21$}                            \\
& & &\\
\textit{GP \texttt{RotationTerm} parameters}            & \multicolumn{3}{l}{} \\
GP rotation period $R_\mathrm{rot,GP}$ [day]            & \multicolumn{3}{l}{$30.63 \pm 3.81$}                       \\
$\sigma_\mathrm{GP}$                                    & \multicolumn{3}{l}{$6.62 \pm 1.45$}                       \\
$Q\mathrm{0}$                                           & \multicolumn{3}{l}{$0.83 \pm 1.48$}                       \\
$dQ$                                                    & \multicolumn{3}{l}{$1.94 \pm 3.67$}                       \\
$f$                                                     & \multicolumn{3}{l}{$0.70 \pm 0.23$}                       \\
& & &\\
Stellar mass $M_\mathrm{s}$ [$M_\mathrm{\odot}$]         & \multicolumn{3}{l}{$ 0.67 \pm 0.06$}      \\
Stellar radius $R_\mathrm{s}$ [$R_\mathrm{\odot}$]         & \multicolumn{3}{l}{$ 0.67 \pm 0.01$}      \\
Stellar density $\rho_\mathrm{s}$ [$\mathrm{g~cm^{-3}}$]         & \multicolumn{3}{l}{$ 3.12 \pm 0.33$}      \\

\hline
\end{tabular}
\end{table*}

\section{Discussion}
\label{sec:discussion}

\begin{figure}
	\includegraphics[width=1\columnwidth]{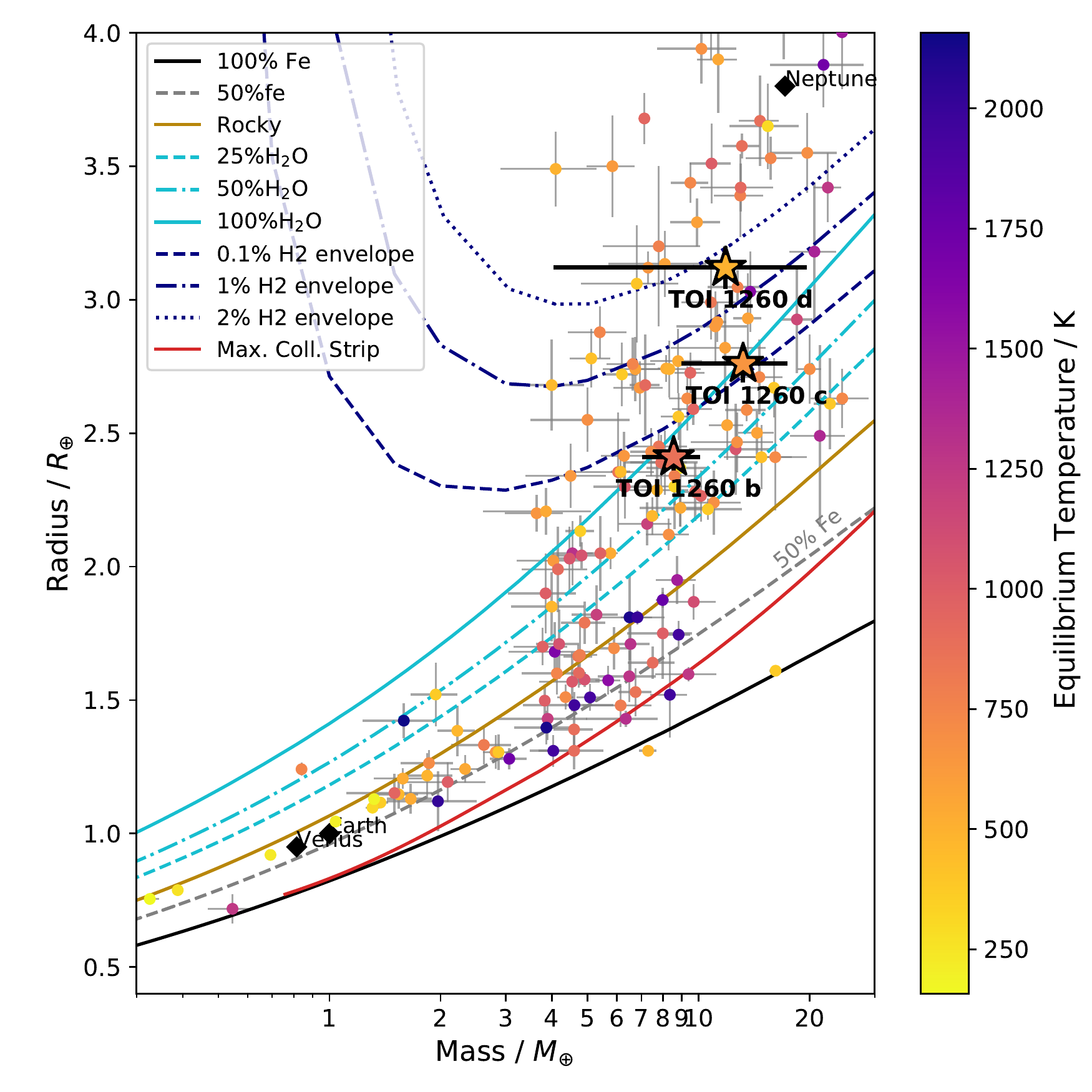}
    \caption{Mass-radius diagram showing low mass planets in the range of 0.4-30 $M_{\oplus}$ which have mass and radius precision measured to better than 30\% and 15\%, respectively.  TOI-1260b, TOI-1260c, TOI-1260d, are indicated by the star symbols. All exoplanets are colour-coded according to their the equilibrium dayside temperatures as shown in the colour bar. The different lines plotted are the theoretical mass-radius relations corresponding to the planet interior compositions \citep{2019PNAS..116.9723Z}.}
    \label{fig:MR_plot}
\end{figure}

The follow-up photometric observations of TOI-1260 allows the precise characterisation of the two inner transiting planets and confirms the planetary nature of the transiting outer planetary companion.
Figure \ref{fig:MR_plot} shows the mass-radius diagram of known exoplanets with masses below 30~$M_{\mathrm{\oplus}}$ and radii less than 4~$R_{\mathrm{\oplus}}$. 
We proceed now with the discussion of the interior composition and atmospheric evolution of the planetary system.

\FloatBarrier

\subsection{Interior composition of the planets}
\label{sec:interior}

The TOI-1260 system has three sub-Neptune transiting exoplanets where planets b, c and d have masses of $8.56 \pm 1.54~M_{\rm \oplus}$, $13.20 \pm 4.23~M_{\rm \oplus}$ and $11.84 \pm 7.79~M_{\rm \oplus}$, respectively, and their radii are $2.41 \pm 0.05~R_{\rm \oplus}$, $2.76 \pm 0.07~R_{\rm \oplus}$, and $3.12 \pm 0.08~R_{\rm \oplus}$, respectively. This means that the three sub-Neptunes TOI-1260 b, c and d have bulk densities of $3.35 \pm 0.64 \rm ~g~cm^{-3}$, $3.45 \pm 1.14 \rm ~g~cm^{-3}$, and $2.14 \pm 1.42 \rm ~g~cm^{-3}$, respectively.
Figure~\ref{fig:MR_plot} shows the distribution of known exoplanet with precise mass and radius measurements in the mass-radius diagram, alongside some theoretical mass-radius relations for different planet interior compositions. The interior of TOI-1260 b is likely to be consisted of up up 50\% rocky core and a 50\% H$_2$O layer. In the case of TOI-1260 c, the sub-Neptune planet is likely a water world or it could be composed of a water-rich core with a small fraction of H2 atmosphere. For the outermost planet TOI-1260 d, its interior is likely to consist of a water-rich or Earth-like rocky core with up to $\sim 2$\% of H2 atmosphere.

The interior compositions of exoplanet correlates with the compositions of their host stars \citep[][]{2021Sci...374..330A}. This is because they were formed from accretion of the same disk material. Therefore, using physical parameters of the host star in addition to the planet's mass and radius provides a better constrain to the planet's interior composition. Using the values of radius, mass, and stellar properties derived in Section \ref{sec:joint_fit}, we performed an analysis of the internal structure of the three planets in the TOI-1260 system. 
Our method is based on a global Bayesian model that fits the observed properties of the star (mass, radius, age, effective temperature, and the photospheric abundances [Si/Fe] and [Mg/Fe]) and planets (planet-star radius ratio, the RV semi-amplitude, and the orbital period). 
The hidden parameters in the Bayesian model are, for each planet, the masses of solids (everything except the H or He gas), the mass fractions of the core, mantle and water, the mass of the gas envelope, the Si/Fe and Mg/Fe mole ratios in the planetary mantle, the S/Fe mole ratio in the core, and the equilibrium temperature. 
All details on the methods are presented in~\citet{Leleu2021}.

The Bayesian analysis relies on a forward models that computes the expected planetary radius and bulk internal structure as a function of the hidden parameters. 
In the forward model, we assume a fully differentiated planet made of a core (composed of Fe and S), a mantle (composed of Si, Mg, Fe, and O), a pure water layer, and a H and He layer. 
The temperature profile is adiabatic, and the equations of state (EoS) used for these calculations are taken from~\citet{2018Icar..313...61H} and~\citet{2016GeoRL..43.6837F} for the core materials, from~\citet{Sotin2007} for the mantle materials, and~\citet{2020A+A...643A.105H} for water. 
The thickness of the gas envelope is determined as a function of the gas mass fraction, the equilibrium temperature, the mass and radius of the solid planet, and the age (assumed to be equal to the stellar age), using the semi-analytical model of~\citet{2014ApJ...792....1L}.
Importantly, the radius of the high-Z part of the planet (core, mantle and water layer) is computed independently of the thickness of the gas layer. 
This implies in particular that the compression effect of the gas envelope onto the core, as well as the effect of the temperature at the basis of the gas envelope are not included in the mode.

The Bayesian analysis is done assuming the following priors: the mass fractions of the planetary cores, mantles, and  water layers have uniform positive priors (the mass fractions of water being limited to a maximum value of 0.5).
The prior on the gas mass is uniform in log, and the bulk Si/Fe and Mg/Fe mole ratios in the planet are assumed to be equal to the values determined for the atmosphere of the star, given above \footnote{It should be noted however that~\citet{Adibekyan2021} has found that despite an existing correlation between the abundances of planets and host stars, the relation is not always strictly one-to-one.}.  

\begin{figure}
     \centering
         \includegraphics[width=\columnwidth]{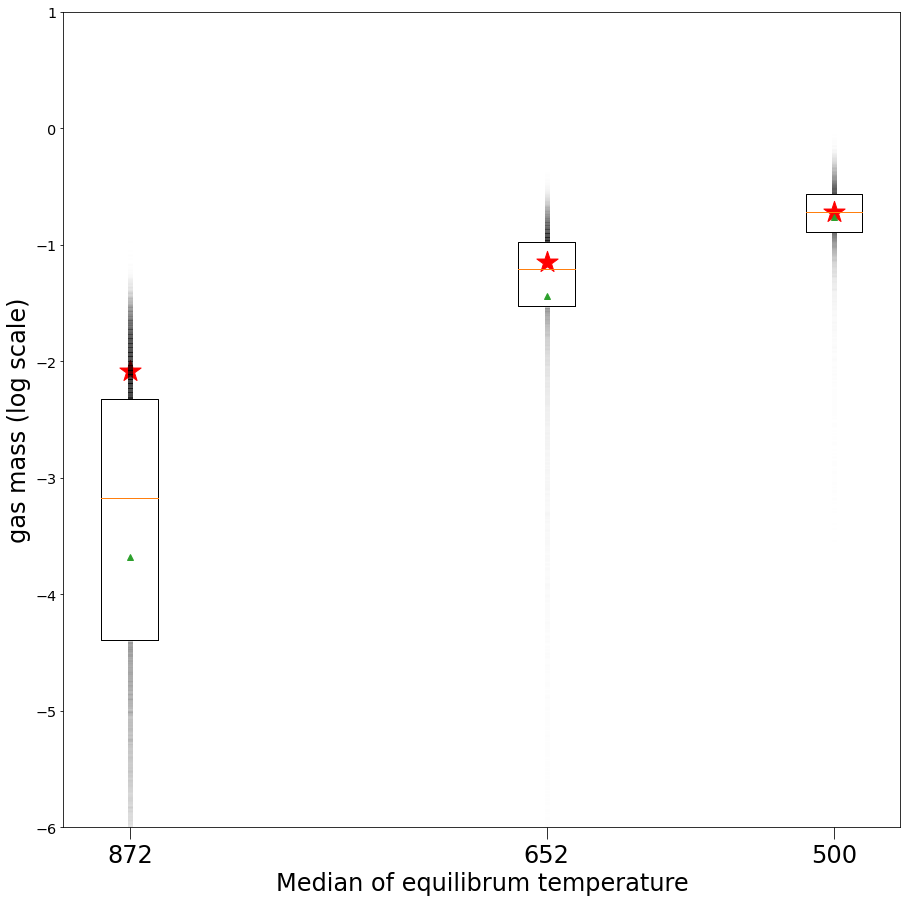}
         \caption{Gas fraction in the planets as a function of their equilibrium temperature. The box show the 25 \% and 75 \% percentiles, the orange line represents the median of the posterior distribution, the green triangle is the mean, and the red stars is located at the mode of the posterior distribution. Finally, the opacity of the thick vertical black line is proportional to the posterior distribution.}
        \label{fig:summary_gas}
\end{figure}

The posterior distribution of the main planetary hidden parameters are presented in Fig.~\ref{fig:corner_IS}. 
All planets have some fraction of gas, the mass of gas increasing for decreasing equilibrium temperatures (see Fig.~\ref{fig:summary_gas}). 
The fraction of water, on the other hand, is essentially unconstrained.  


\begin{figure*}
     \centering
     \begin{subfigure}[b]{1.3\columnwidth}
         \centering
         \includegraphics[width=\columnwidth]{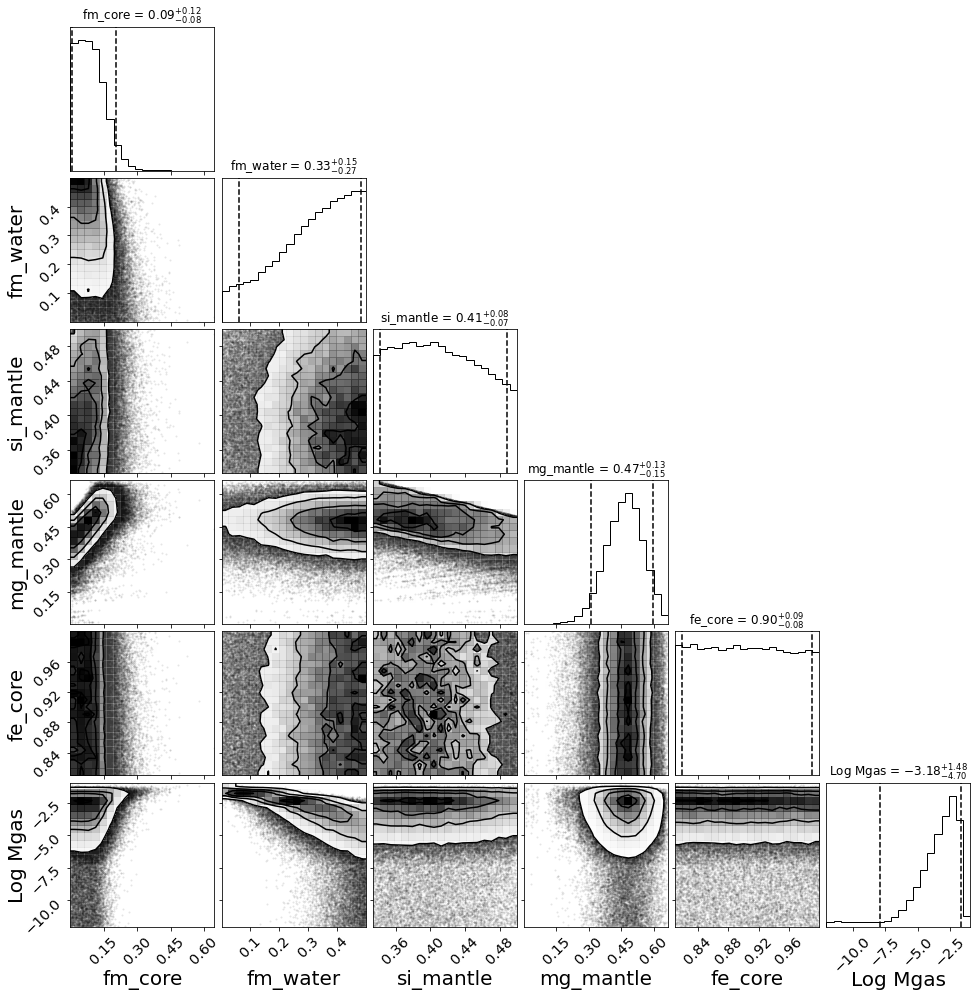}
         \subcaption{}
     \end{subfigure}
     \hfill
     \begin{subfigure}[b]{1.3\columnwidth}
         \centering
         \includegraphics[width=\columnwidth]{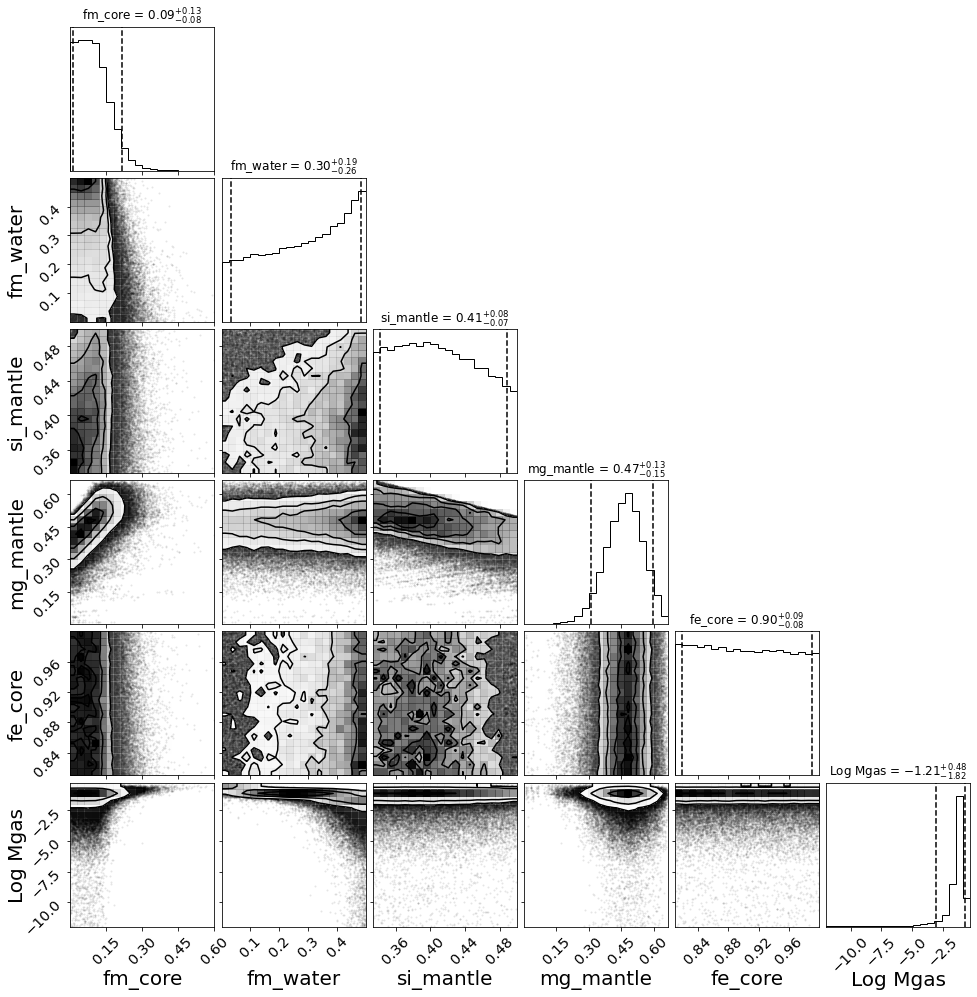}
         \subcaption{}
     \end{subfigure}
\end{figure*}

\begin{figure*}\ContinuedFloat
    \centering
    \begin{subfigure}[b]{1.3\columnwidth}
        \includegraphics[width=\columnwidth]{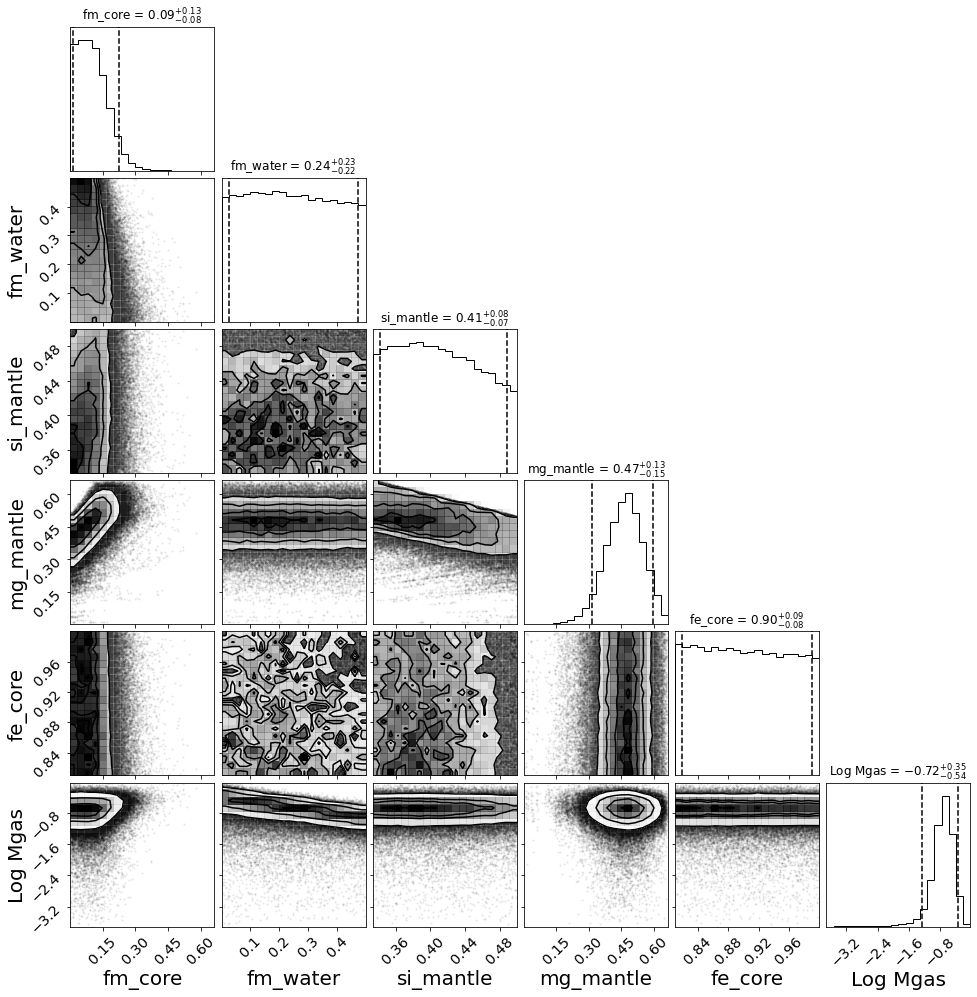}
        \subcaption{}
    \end{subfigure}
    \caption{Corner plot showing the results on the interior composition models of (a) TOI-1260 b, (b) TOI-1260 c and (c) TOI-1260 d. The vertical dashed lines and the 'error bars' given at the top of each columns represent the 5 \% and 95 \% percentiles.}
    \label{fig:corner_IS}
\end{figure*}


\subsection{Atmospheric evolution}

We considered the stellar and planetary parameters derived in our paper, as well as the present-day planetary atmospheric mass fractions presented in Section~\ref{sec:interior}, to reconstruct the evolution of the stellar rotation rate and of the planetary atmospheres. 
In particular, we constrain the evolution of the stellar rotation period, which we use as proxy for the evolution of the stellar high-energy emission affecting atmospheric escape, and the predicted initial atmospheric mass fraction of the detected planets $f_{\rm atm}^{\rm start}$, that is the mass of the planetary atmosphere at the time of the dispersal of the protoplanetary disk, which we assume being at 5\,Myr. 

We reach these results by using the \textit{P}lanetary \textit{A}tmospheres and \textit{S}tellar Ro\textit{T}ation R\textit{A}tes~\citep[PASTA;][]{bonfanti2021_atmo} code, which is an updated version of the original code presented by~\citet{kubyshkina2019a,kubyshkina2019b}. 
In short, PASTA constraints the evolution of planetary atmospheres and of the stellar rotation rate combining a model predicting planetary atmospheric escape rates based on hydrodynamic simulations~\citep[this has the advantage over other commonly used analytical estimates to account for both XUV-driven and core-powered mass loss;][]{kubyshkina2018}, a model of the stellar high-energy (X-ray plus extreme ultraviolet; XUV) flux evolution \citep{bonfanti2021_atmo}, a model relating planetary parameters and atmospheric mass~\citep{johnstone15models}, and stellar evolutionary tracks~\citep{Choi2016}. 
PASTA works under two main assumptions: 1) planet migration did not occur after the dispersal of the protoplanetary disk; 2) the planets hosted at some point in the past or still host a hydrogen-dominated atmosphere. 
PASTA returns realistic uncertainties on the free parameters (i.e. the planetary initial atmospheric mass fractions at the time of the dispersal of the protoplanetary disk, and the indexes of the power law controlling the stellar rotation period that is used as proxy for the stellar XUV emission) by implementing the atmospheric evolution algorithm in a Bayesian framework~\citep{cubillos2017}, using the system parameters with their uncertainties as input priors. 
All details of the algorithm can be found in~\citet{bonfanti2021_atmo}. 
The only difference with respect to the analysis of the systems considered by~\citet{bonfanti2021_atmo} is that here we fit the planetary atmospheric mass fractions given in Section~\ref{sec:interior} instead of the planetary radii.
This enables the code to be more accurate by avoiding the continuous conversion of the atmospheric mass fraction into planetary radius, given the other system parameters~\citep[see e.g.][]{2021NatAs...5..775D}.

\begin{figure*}
	\includegraphics[width=2\columnwidth]{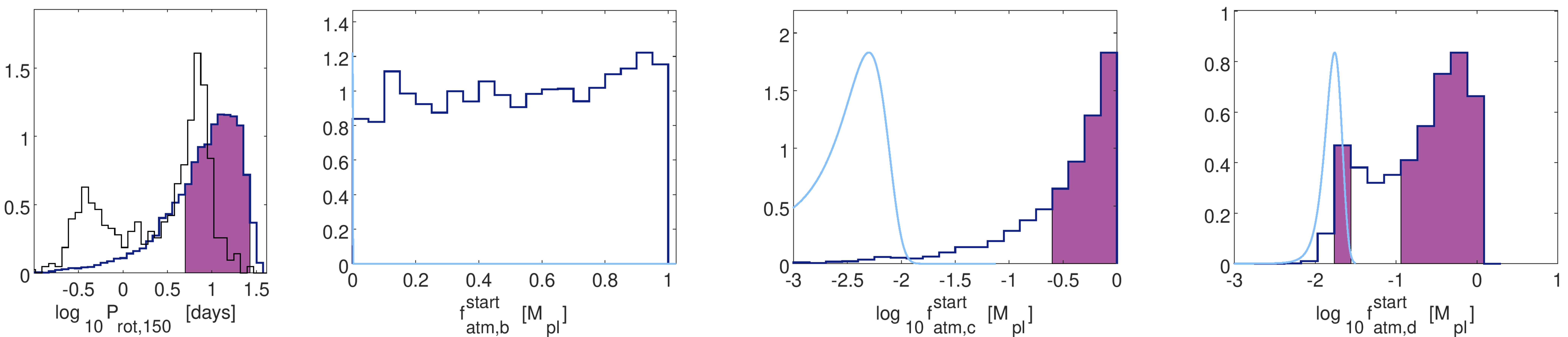}
    \caption{Posterior probability density functions (PDFs) of the stellar rotation period when TOI-1260 was 150 Myr old ($P_{\mathrm{rot,150}}$) and of the initial atmospheric mass fraction ($f_{\mathrm{atm}}^{\mathrm{start}}$) of the hosted exoplanets. The purple areas show the 68\%-HPD (highest posterior density) interval. \textit{Leftmost panel.} $P_{\mathrm{rot,150}}$ PDF (dark blue histogram) to be compared with the rotation period distribution of stars of comparable masses that belong to coeval open clusters~\citep[black histogram; data taken from][]{johnstone15Prot150}. \textit{Other panels.} Atmospheric mass fractions PDFs of planet b (linear scale) and of planet c and d (log scale). The light blue curve is the present-day atmospheric content, as inferred from our internal structure analysis. See text for details. \label{fig:atmos_evolution}}
\end{figure*}

Figure~\ref{fig:atmos_evolution} shows the results obtained from PASTA. 
As a proxy for the evolution of the stellar rotation period, Figure~\ref{fig:atmos_evolution} displays the posterior distribution of the stellar rotation period at an age of 150\,Myr ($P_{\mathrm{rot,150}}$), also in comparison to that of stars member of young open clusters and of comparable mass extracted from~\citet{johnstone15Prot150}. 
The posterior distribution is slightly shifted towards slower rotation compared to that of the open cluster stars, indicating that the planets were likely subject to somewhat less XUV radiation than the average. 

Figure~\ref{fig:atmos_evolution} shows also the posterior distribution of the initial atmospheric mass fraction for planets b (in linear scale), c (in logarithmic scale), and d (in logarithmic scale) in comparison to the present-day atmospheric mass fraction (Section~\ref{sec:interior}). 
The posterior distribution for planet b is flat, indicating that the planet has most likely lost (almost) entirely its primordial hydrogen-dominated envelope through escape at some point in the past, which is why PASTA is unable to constrain the initial atmospheric mass fraction. Figure~\ref{fig:atmos_evolution} indicates that also planets c and d have gone through significant evolution through escape that has significantly eroded the primordial atmospheric content, which was however small in comparison to the planetary masses. 
Therefore, we conclude that both planets (i.e. c and d) accreted a small hydrogen envelope during the formation process compared to their masses. 
This may have been the result of several physical mechanisms, such as late planet formation compared to the age of the protoplanetary disk, early dispersal of the protoplanetary disk, low gas content of the disk.

As the isochronal age is loosely constrained, we performed additional evolution runs by artificially making the star much younger or older, further imposing tighter constraints on the stellar age. Despite the different evolutionary time scales, we did not find significant changes in the $f_{\mathrm{atm}}^{\mathrm{start}}$ of the planets. This is because (1) atmospheric mass loss is significant only during the first Myrs of evolution and (2) $f_{\mathrm{atm,c}}^{\mathrm{start}}$ and $f_{\mathrm{atm,d}}^{\mathrm{start}}$ are always found to be rather small, indicating that the constraints given by system parameters prevent those planets to host a massive initial atmosphere regardless of the age of the system. 

The current stellar XUV fluxes impinging on each planet are $F_\mathrm{XUV,b} = 2.87\cdot10^{4}$ erg/(cm$^2$ s), $F_\mathrm{XUV,c} = 8.97\cdot10^3$ erg/(cm$^2$ s), and $F_\mathrm{XUV,d} = 3.10\cdot10^3$ erg/(cm$^2$ s).
The correspondent mass-loss rate values expected for the planets right now are $\dot{M}_b = 10^{10}$ g/s, $\dot{M}_c = 1.59\cdot10^9$ g/s, and $\dot{M}_d = 7.43\cdot10^8$ g/s.
Assuming that the stellar XUV flux does not change over time in the future, which is a reasonable assumption given the old age of the star, these values imply that in the next Gyr the planets are respectively going to lose 0.6\%, 0.06\%, and 0.03\% of their mass.
From Fig.~\ref{fig:summary_gas} these values then imply that planet b is going to lose entirely its hydrogen-dominated envelope, while planets c and d are going to keep it.
As the results of planet b are consistent with no hydrogen atmosphere at all, it is unlikely that the position of these planets in the period-radius diagram~\citep[e.g.][]{2017AJ....154..109F} is going to change in the future.

\section{Conclusions}
\label{sec:conclusion}

We presented the follow-up observations of the TOI-1260 system using CHEOPS and TESS. 
The addition of the recent photometric dataset allow us to refine the physical parameters of the planetary system and discover a third additional transiting planet. 
For planets TOI-1260 b and c, we found that the radii are $2.36 \pm 0.06 \rm ~R_{\oplus}$, $2.82 \pm 0.08 \rm ~R_{\oplus}$, respectively, and the masses $8.52 \pm 1.45 \rm ~M_{\oplus}$ and $13.29 \pm 3.94 \rm ~M_{\oplus}$ . 
The newly discovered TOI-1260-d has bulk properties $3.01 \pm 0.09 \rm ~R_{\oplus}$ and $11.8 \pm 7.5 \rm ~M_{\oplus}$. 

The detailed characterization of the planetary parameters allows us to derive constraints of their internal composition and evolution that we related to the formation processes in the system and its future evolution.

\begin{figure}
    \includegraphics[width=1\columnwidth]{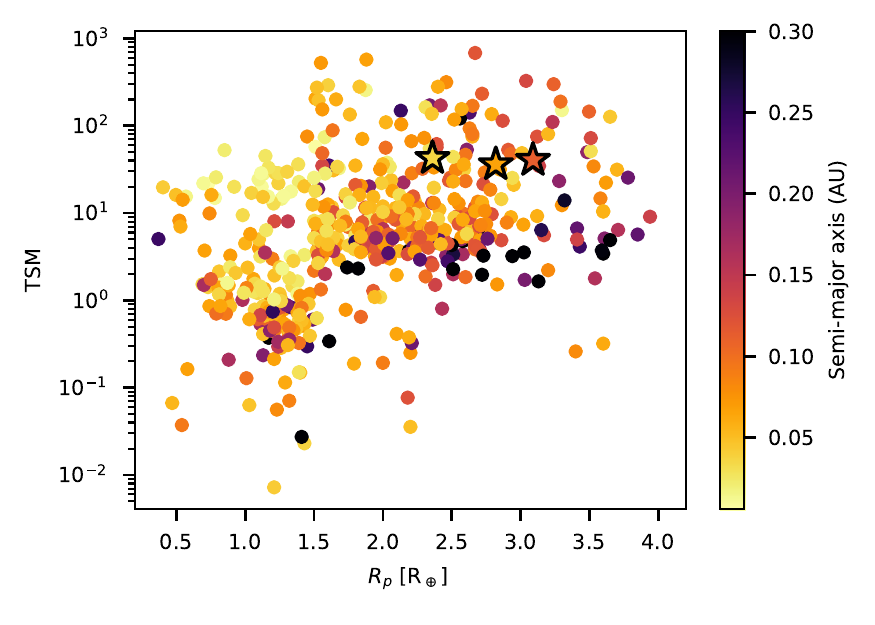}
    \caption{Unscaled transmission spectroscopy metrics (TSM) of all multi-planet systems with host stars of K-type and later as a function of planetary radius, with orbital separation visible in the colour scale. The three planets in the TOI-1260 system are marked with stars.}
    \label{fig:TSMs}
\end{figure}

The TOI-1260 system presents an exciting opportunity for comparative exoplanetology using JWST transmission spectroscopy. 
\citet{Moses2013} predicted that sub-Neptune sized exoplanets such as those in the TOI-1260 system can harbour a large diversity of atmospheric compositions. 
Multi-planet systems such as TOI-1260 give us the opportunity to test whether such diversity can exist within different sub-Neptunes in the same system. 
All three of the planets in the TOI-1260 system appear to be favourable for atmospheric categorisation with JWST, with transmission spectroscopy metrics \citep[TSMs][]{Kempton2018} of 43.6, 36.1 and 40.4 for planets b, c, and d, respectively. 
Figure \ref{fig:TSMs} shows how this compares to similar multi-planet systems as a function of planetary radius and semi-major axis. 
In addition, due to its high northern declination TOI-1260 is particularly favourable for JWST visibility, with observations possible for 196 days each year \citep{exoctk}.

\FloatBarrier


\section*{Acknowledgements}


CHEOPS is an ESA mission in partnership with Switzerland with important contributions to
the payload and the ground segment from Austria, Belgium, France, Germany, Hungary, Italy,
Portugal, Spain, Sweden, and the United Kingdom. The CHEOPS Consortium would like to
gratefully acknowledge the support received by all the agencies, offices, universities, and
industries involved. Their flexibility and willingness to explore new approaches were essential
to the success of this mission.\\
KGI is the ESA CHEOPS Project Scientist and is responsible for the ESA CHEOPS Guest
Observers Programme. She does not participate in, or contribute to, the definition of the
Guaranteed Time Programme of the CHEOPS mission through which observations described
in this paper have been taken, nor to any aspect of target selection for the programme.\\
K.W.F.L. acknowledge support by DFG grants RA714/14-1 within the DFG Schwerpunkt SPP 1992, “Exploring the Diversity of Extrasolar Planets”.\\
YA, MJH and JAE acknowledge  the  support  of  the  Swiss  National  Fund  under  grant 200020\_172746.\\
CMP, MF and IYG gratefully acknowledge the support of the Swedish National Space Agency (DNR 65/19, 174/19, 174/18).\\
SH gratefully acknowledges CNES funding through the grant 837319.\\
ACC and TW acknowledge support from STFC consolidated grant numbers ST/R000824/1 and ST/V000861/1, and UKSA grant number ST/R003203/1.\\
We acknowledge support from the Spanish Ministry of Science and Innovation and the European Regional Development Fund through grants ESP2016-80435-C2-1-R, ESP2016-80435-C2-2-R, PGC2018-098153-B-C33, PGC2018-098153-B-C31, ESP2017-87676-C5-1-R, MDM-2017-0737 Unidad de Excelencia ''Mar{\'i}a de Maeztu''- Centro de Astrobiolog{'i}a (INTA-CSIC), as well as the support of the Generalitat de Catalunya/CERCA programme. 
The MOC activities have been supported by the ESA contract No. 4000124370.\\
S.C.C.B. acknowledges support from FCT through FCT contracts nr. IF/01312/2014/CP1215/CT0004.\\
XB, SC, DG, MF and JL acknowledge their role as ESA-appointed CHEOPS science team members.\\
ABr was supported by the SNSA.\\
This project was supported by the CNES.\\
LD is an F.R.S.-FNRS Postdoctoral Researcher. The Belgian participation to CHEOPS has been supported by the Belgian Federal Science Policy Office (BELSPO) in the framework of the PRODEX Program, and by the University of Li{\`e}ge through an ARC grant for Concerted Research Actions financed by the Wallonia-Brussels Federation.\\
This work was supported by FCT - Funda\c{c}{\~a}o para a Ci{\^e}ncia e a Tecnologia through national funds and by FEDER through COMPETE2020 - Programa Operacional Competitividade e Internacionaliza\c{c}{\~a}o by these grants: UID/FIS/04434/2019; UIDB/04434/2020; UIDP/04434/2020; PTDC/FIS-AST/32113/2017 \& POCI-01-0145-FEDER-032113; PTDC/FIS-AST/28953/2017 \& POCI-01-0145-FEDER-028953; PTDC/FIS-AST/28987/2017 \& POCI-01-0145-FEDER-028987. 
O.D.S.D. is supported in the form of work contract (DL 57/2016/CP1364/CT0004) funded by national funds through FCT.\\
B.-O.D. acknowledges support from the Swiss National Science Foundation (PP00P2-190080).\\
This project has received funding from the European Research Council (ERC) under the European Union’s Horizon 2020 research and innovation programme (project {\sc Four Aces}; grant agreement No 724427). It has also been carried out in the frame of the National Centre for Competence in Research PlanetS supported by the Swiss National Science Foundation (SNSF). DE acknowledges financial support from the Swiss National Science Foundation for project 200021\_200726.\\
DG gratefully acknowledges financial support from the CRT foundation under Grant No. 2018.2323 ''Gaseousor rocky? Unveiling the nature of small worlds''.\\
M.G. is an F.R.S.-FNRS Senior Research Associate.\\
This work was granted access to the HPC resources of MesoPSL financed by the Region Ile de France and the project Equip@Meso (reference ANR-10-EQPX-29-01) of the programme Investissements d'Avenir supervised by the Agence Nationale pour la Recherche.\\
ML acknowledges support of the Swiss National Science Foundation under grant number PCEFP2\_194576.\\
GSc, GPi, IPa, LBo, VNa and RRa acknowledge the funding support from Italian Space Agency (ASI) regulated by "Accordo ASI-INAF n. 2013-016-R.0 del 9 luglio 2013 e integrazione del 9 luglio 2015 CHEOPS Fasi A/B/C".\\
PM acknowledges support from STFC research grant number ST/M001040/1.\\
This work was also partially supported by a grant from the Simons Foundation (PI Queloz, grant number 327127).\\
IR acknowledges support from the Spanish Ministry of Science and Innovation and the European Regional Development Fund through grant PGC2018-098153-B- C33, as well as the support of the Generalitat de Catalunya/CERCA programme.\\
S.G.S. acknowledge support from FCT through FCT contract nr. CEECIND/00826/2018 and POPH/FSE (EC).\\
GyMSz acknowledges the support of the Hungarian National Research, Development and Innovation Office (NKFIH) grant K-125015, a PRODEX Institute Agreement between the ELTE E\"otv\"os Lor\'and University and the European Space Agency (ESA-D/SCI-LE-2021-0025), the Lend\"ulet LP2018-7/2021 grant of the Hungarian Academy of Science and the support of the city of Szombathely.\\
V.V.G. is an F.R.S-FNRS Research Associate. \\
S.S. has received funding from the European Research Council (ERC) under the European Union’s Horizon 2020 research and innovation programme (grant agreement No 833925, project STAREX).
R.L. acknowledges funding from University of La Laguna through the Margarita Salas Fellowship from the Spanish Ministry of Universities ref. UNI/551/2021-May 26, and under the EU Next Generation funds.
This paper includes data collected with the TESS mission, obtained from the MAST data archive at the Space Telescope Science Institute (STScI). Funding for the TESS mission was provided by the NASA Explorer Program. STScI is operated by the Association of Universities for Research in Astronomy, Inc., under NASA contract NAS 5–26555.

\section*{Data Availability}

TESS data are publicly available in the Space Telescope Science Institute (STScI) at \url{https://mast.stsci.edu}.
CHEOPS data generated and analysed in this article will be made available in the CHEOPS mission archive (\url{https://cheops.unige.ch/archive_browser/}).
No new HARPS RV data were generated for this work.



\bibliographystyle{mnras}
\bibliography{toi-1260_cheops} 

\bigskip
\bigskip


$^{1}$~Institute of Planetary Research, German Aerospace Center (DLR), Rutherfordstrasse 2, 12489 Berlin, Germany\\
$^{2}$~Physikalisches Institut, University of Bern, Sidlerstrasse 5, 3012 Bern, Switzerland\\
$^{3}$~Cavendish Laboratory, JJ Thomson Avenue, Cambridge CB3 0HE, UK\\
$^{4}$~Space Research Institute, Austrian Academy of Sciences, Schmiedlstrasse 6, A-8042 Graz, Austria\\
$^{5}$~Observatoire Astronomique de l'Universit\'e de Gen\`eve, Chemin Pegasi 51, Versoix, Switzerland\\
$^{6}$~Department of Astronomy, Stockholm University, AlbaNova University Center, 10691 Stockholm, Sweden; 
$^{7}$~Department of Space, Earth and Environment, Chalmers University of Technology, Onsala Space Observatory, 43992 Onsala, Sweden\\
$^{8}$~Leiden Observatory, University of Leiden, PO Box 9513, 2300 RA Leiden, The Netherlands\\
$^{9}$~Aix Marseille Univ, CNRS, CNES, LAM, 38 rue Fr{'e}d{'e}ric Joliot-Curie, 13388 Marseille, France\\
$^{10}$~Center for Space and Habitability, University of Bern, Gesellschaftsstrasse 6, 3012 Bern, Switzerland\\
$^{11}$~Department of Physics and Kavli Institute for Astrophysics and Space Research, Massachusetts Institute of Technology, Cambridge, MA 02139, USA\\
$^{12}$~Centre for Exoplanet Science, SUPA School of Physics and Astronomy, University of St Andrews, North Haugh, St Andrews KY16 9SS, UK\\
$^{13}$~Instituto de Astrof{\'i}sica de Canarias, 38200 La Laguna, Tenerife, Spain\\
$^{14}$~Departamento de Astrof{\'i}sica, Universidad de La Laguna, 38206 La Laguna, Tenerife, Spain\\
$^{15}$~Instituto de Astrof{\'i}sica e Ci{\^e}ncias do Espa\c{c}o, Universidade do Porto, CAUP, Rua das Estrelas, 4150-762 Porto, Portugal\\
$^{16}$~Departamento de F{\'i}sica e Astronomia, Faculdade de Ci{\^e}ncias, Universidade do Porto, Rua do Campo Alegre, 4169-007 Porto, Portugal\\
$^{17}$~Institut de Ci{\`e}ncies de l'Espai (ICE, CSIC), Campus UAB, Can Magrans s/n, 08193 Bellaterra, Spain\\
$^{18}$~Institut d'Estudis Espacials de Catalunya (IEEC), 08034 Barcelona, Spain\\
$^{19}$~Admatis, 5. Kand{'o} K{'a}lm{'a}n Street, 3534 Miskolc, Hungary\\
$^{20}$~Depto. de Astrofísica, Centro de Astrobiolog{\'i}a (CSIC-INTA), ESAC campus, 28692 Villanueva de la Ca{\~n}ada (Madrid), Spain\\
$^{21}$~Universit{\'e} Grenoble Alpes, CNRS, IPAG, 38000 Grenoble, France\\
$^{22}$~Universit{\'e} de Paris, Institut de physique du globe de Paris, CNRS, F-75005 Paris, France\\
$^{23}$~Lund Observatory, Dept. of Astronomy and Theoretical Physics, Lund University, Box 43, 22100 Lund, Sweden\\
$^{24}$~Astrobiology Research Unit, Universit{\'e} de Li{\`e}ge, All{\'e}e du 6 Ao{\^u}t 19C, B-4000 Li{\`e}ge, Belgium\\
$^{25}$~Space sciences, Technologies and Astrophysics Research (STAR) Institute, Universit{\'e} de Li{\`e}ge, All{\'e}e du 6 Ao{\^u}t 19C, 4000 Li{\`e}ge, Belgium\\
$^{26}$~Dipartimento di Fisica, Universit{\`a} degli Studi di Torino, via Pietro Giuria 1, I-10125, Torino, Italy\\
$^{27}$~Department of Astrophysics, University of Vienna, Tuerkenschanzstrasse 17, 1180 Vienna, Austria\\
$^{28}$~Division Technique INSU, BP 330, 83507 La Seyne cedex, France\\
$^{29}$~IMCCE, UMR8028 CNRS, Observatoire de Paris, PSL Univ., Sorbonne Univ., 77 av. Denfert-Rochereau, 75014 Paris, France\\
$^{30}$~Center for Astrophysics, Harvard \& Smithsonian, 60 Garden Street, Cambridge, MA 02138, USA\\
$^{31}$~Institut d'astrophysique de Paris, UMR7095 CNRS, Universit{'e} Pierre \& Marie Curie, 98bis blvd. Arago, 75014 Paris, France\\
$^{32}$~Department of Physics, University of Warwick, Gibbet Hill Road, Coventry CV4 7AL, United Kingdom\\
$^{33}$~Science and Operations Department - Science Division (SCI-SC), Directorate of Science, European Space Agency (ESA), European Space Research and Technology Centre (ESTEC), Keplerlaan 1, 2201-AZ Noordwijk, The Netherlands\\
$^{34}$~Konkoly Observatory, Research Centre for Astronomy and Earth Sciences, 1121 Budapest, Konkoly Thege Mikl{\'o}s {\'u}t 15-17, Hungary\\
$^{35}$~ELTE E{\"o}tv{\"o}s Lor{\'a}nd University, Institute of Physics, P{\'a}zm{\'a}ny P{\'e}ter s{\'e}t{\'a}ny 1/A, 1117 Budapest, Hungary\\
$^{36}$~Sydney Institute for Astronomy, School of Physics A29, University of Sydney, NSW 2006, Australia\\
$^{37}$~INAF, Osservatorio Astronomico di Padova, Vicolo dell'Osservatorio 5, 35122 Padova, Italy\\
$^{38}$~Astrophysics Group, Keele University, Staffordshire, ST5 5BG, United Kingdom\\
$^{39}$~INAF, Osservatorio Astrofisico di Catania, Via S. Sofia 78, 95123 Catania, Italy\\
$^{40}$~Institute of Optical Sensor Systems, German Aerospace Center (DLR), Rutherfordstrasse 2, 12489 Berlin, Germany\\
$^{41}$~ETH Zurich, Department of Physics, Wolfgang-Pauli-Strasse 2, CH-8093 Zurich, Switzerland\\
$^{42}$~Dipartimento di Fisica e Astronomia "Galileo Galilei", Universita degli Studi di Padova, Vicolo dell'Osservatorio 3, 35122 Padova, Italy
$^{43}$~ESTEC, European Space Agency, 2201AZ, Noordwijk, NL\\
$^{44}$~Zentrum f{\"ur} Astronomie und Astrophysik, Technische Universit{\"a}t Berlin, Hardenbergstr. 36, D-10623 Berlin, Germany\\
$^{45}$~Institut für Geologische Wissenschaften, Freie Universit{\"a}t Berlin, 12249 Berlin, Germany\\
$^{46}$~Department of Earth, Atmospheric and Planetary Sciences, Massachusetts Institute of Technology, Cambridge, MA 02139, USA\\
$^{47}$~Department of Aeronautics and Astronautics, MIT, 77 Massachusetts Avenue, Cambridge, MA 02139, USA\\
$^{48}$~MTA-ELTE Exoplanet Research Group, 9700 Szombathely, Szent Imre h. u. 112, Hungary\\
$^{49}$~Institute of Astronomy, University of Cambridge, Madingley Road, Cambridge, CB3 0HA, United Kingdom\\
$^{50}$~Department of Astrophysical Sciences, Princeton University, 4 Ivy Ln, Princeton, NJ 08544, USA\\
$^{51}$~Department of Astronomy \& Astrophysics, University of Chicago, Chicago, IL 60637, USA\\




\appendix

\section{Extra material}

\begin{figure*}
    \centering
	\includegraphics[width=2.25\columnwidth]{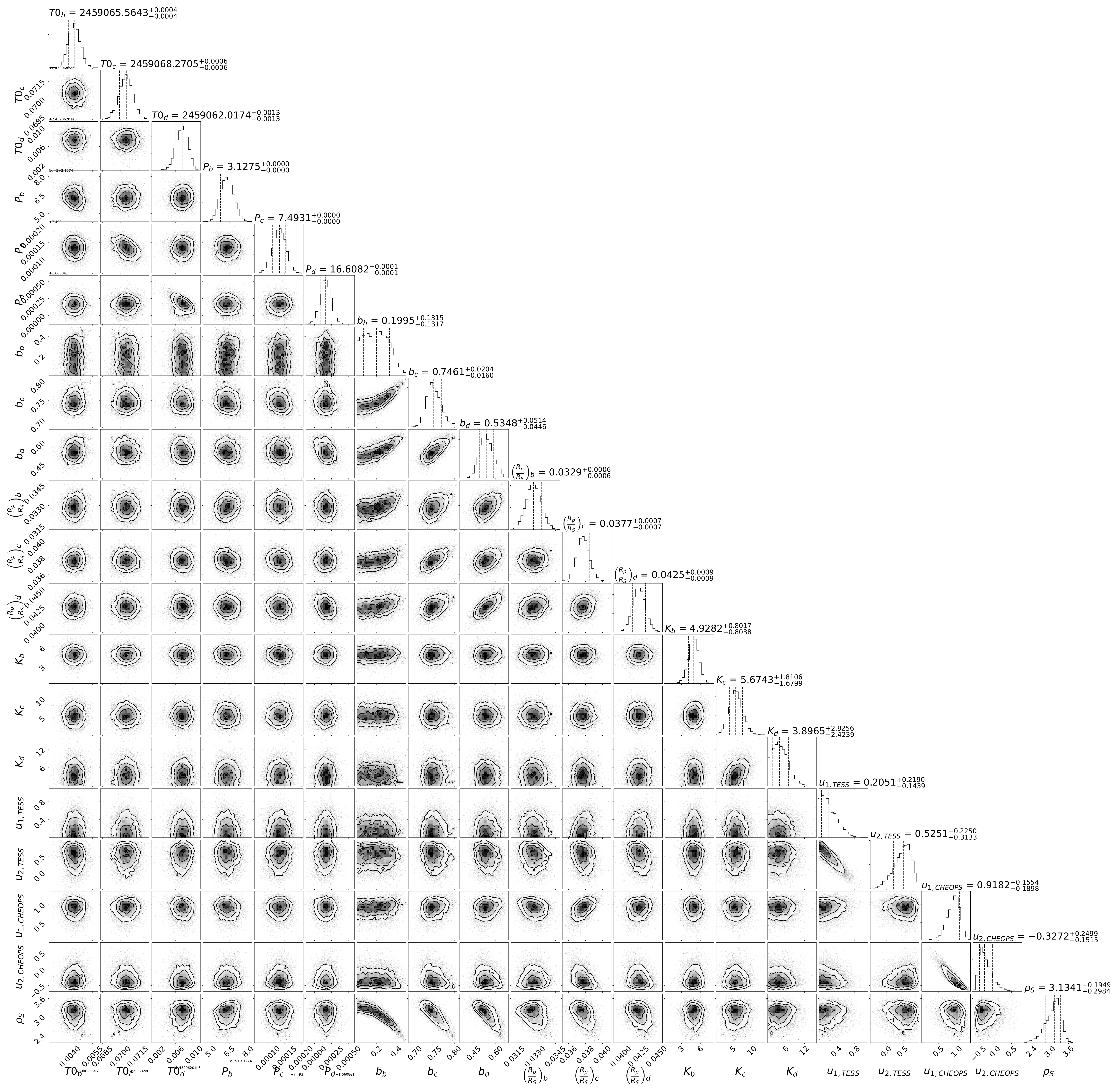}
    \caption{Corner plot showing the posterior distribution of the fitted parameters.}
    \label{fig:corner_plot}
\end{figure*}

We present in Figure~\ref{fig:corner_plot} the posterior distribution of the fitted parameters.

\bsp	
\label{lastpage}
\end{document}